\documentclass[prd,twocolumn,amsmath,amssymb,floatfix,superscriptaddress,nofootinbib]{revtex4-1}

\usepackage{natbib}
\usepackage{amsmath}
\usepackage{amssymb}
\usepackage{latexsym}
\usepackage{bm}   
\usepackage{graphicx,epstopdf, epsfig}
\usepackage{color}
\usepackage{hyperref}
\DeclareMathAlphabet\mathbfcal{OMS}{cmsy}{b}{n}
\definecolor{darkgreen}{cmyk}{0.85,0.2,1.00,0.35} 
\definecolor{purple}{cmyk}{0.5,1.0,0,0} 
\definecolor{darkblue}{cmyk}{1.0,1.0,0,0}
\hypersetup{
    colorlinks = true,
    allcolors = darkblue
}

\newcommand{\be}{\begin{equation}}
\newcommand{\ee}{\end{equation}}
\newcommand{\curv}{{\cal R}}
\newcommand{\alphap}{\alpha^+}
\newcommand{\alpham}{\alpha^-}
\newcommand{\fs}{f_*}
\newcommand{\bogo}{Bogoliubov }  
\newcommand{\NL}{{\rm NL}}
\newcommand{\bp}{{\chi}}


\begin{document}

\preprint{}

\title{Nonlinear Excitations in Inflationary Power Spectra}

\author{Vinicius  Miranda}
\affiliation{Center for Particle Cosmology, Department of Physics and Astronomy, University of Pennsylvania, Philadelphia, PA 19104, U.S.A.}
\affiliation{Kavli Institute for Cosmological Physics, The University of Chicago, Chicago, Illinois 60637, U.S.A.}
\affiliation{Department of Astronomy \& Astrophysics, University of Chicago, Chicago IL 60637, U.S.A.}

\author{Wayne Hu}
\affiliation{Kavli Institute for Cosmological Physics, The University of Chicago, Chicago, Illinois 60637, U.S.A.}
\affiliation{Department of Astronomy \& Astrophysics, University of Chicago, Chicago IL 60637, U.S.A.}

\author{Chen He}
\affiliation{Kavli Institute for Cosmological Physics, The University of Chicago, Chicago, Illinois 60637, U.S.A.}
\affiliation{Department of Astronomy \& Astrophysics, University of Chicago, Chicago IL 60637, U.S.A.}

\author{Hayato Motohashi}
\affiliation{Kavli Institute for Cosmological Physics, The University of Chicago, Chicago, Illinois 60637, U.S.A.}

\begin{abstract}%%%%%%%%%%%%%%%%%%%%%%%%%%%%%%%%%%%%%%%%%
We develop methods to calculate the  curvature power spectrum  in models where  features in the inflaton potential nonlinearly excite modes   and generate  high frequency features in the spectrum.
%Due to projection effects in the CMB, a large amplitude high frequency feature produces much smaller
%effects in the temperature power spectrum. 
The first nontrivial effect of  excitations generating further excitations 
arises at third order in deviations from slow roll.    If these further excitations are contemporaneous, the series can
be resummed, showing the exponential sensitivity of the curvature spectrum to potential features.
More generally, this exponential approximation
provides a power spectrum template which nonlinearly obeys relations between 
excitation coefficients and whose parameters may be appropriately
adjusted.   For a large sharp step in the potential, it greatly improves  the analytic power spectrum template and its
dependence on potential parameters.   For axionic oscillations in the potential, it
corrects the mapping between the potential and the amplitude, phase and zero point of the 
curvature oscillations, which might otherwise cause erroneous inferences in for example the tensor-scalar ratio,
formally even when that amplitude is $10^3$ times larger than the slow roll power spectrum.
It also estimates when terms that produce double frequency oscillations
that are usually omitted when analyzing data should be included.
These techniques should allow future studies of high frequency features in
the CMB and large scale structure to extend to higher amplitude and/or higher precision.  
\end{abstract}


\date{\today}

\maketitle

\section{Introduction}%%%%%%%%%%%%%%%%%%%%%%%%%%%%%%%%%%%%%%%%%

Features in the inflaton potential produce features in the curvature spectrum that are imprinted into
cosmological observables of the cosmic microwave background (CMB) and large scale structure.
In particular, the CMB power spectrum places strong, model-independent constraints
on the amplitude of broad features that persist over more than an efold in wavenumber 
(e.g.~\cite{Hannestad:2000pm,Hu:2003vp,Tegmark:2002cy,Hannestad:2003zs,Bridle:2003sa,Mukherjee:2003ag,Leach:2005av,Peiris:2009wp,Hlozek:2011pc,Gauthier:2012aq,Vazquez:2012ux,Hunt:2013bha,Aslanyan:2014mqa,Hazra:2014jwa,Ade:2015lrj}).

Sharp or high frequency features are more difficult to constrain both because
they violate the ordinary slow roll approximation and their observable impact in the CMB is
suppressed due to projection effects.  Such features also have the ability to mimic
local statistical fluctuations in data (e.g.~\cite{Meerburg:2013cla,Easther:2013kla}) and so require accurate
model predictions over a range of scales or observables to detect.   This is often undertaken on 
a case by case basis, for example in axion monodromy \cite{Silverstein:2008sg} 
and step potential models \cite{Adams:2001vc}.  Indeed predictions for
these two models have been extensively developed
(e.g.~\cite{Miranda:2013wxa,Flauger:2014ana} for improvements in range
required for Planck CMB data)
due to their ability to fit various anomalies
in the CMB power spectrum \cite{Peiris:2003ff,Flauger:2009ab}.

In this paper, we develop general techniques for predicting the curvature
power spectrum for models with large, high frequency features.   These features
necessitate an extension of the ordinary slow roll approximation where not all slow roll parameters
are considered to be slowly varying.  Instead the prediction for the inflaton modefunction
excitations can be  iteratively improved using Green function techniques   \cite{Stewart:2001cd}, typically up to at most
second order in deviations from slow roll \cite{Choe:2004zg}.

In Ref.~\cite{Dvorkin:2009ne} these techniques were further developed to predict  percent level
accurate CMB temperature power spectra for nearly order unity curvature fluctuations. 
By requiring conservation of
superhorizon curvature fluctuations and positive definite power spectra, these techniques
also provide a controlled approximation for the amplitude of even larger features but with diminished accuracy for their form.
Analyses that utilized this technique imposed a prior on the amplitude of features
that were not necessarily required by the data \cite{Dvorkin:2010dn,Dvorkin:2011ui,Miranda:2014fwa}.   Furthermore
the Planck data now require sub-percent level accuracy for small-scale features.   
Here we present techniques that go beyond second order perturbation theory to describe the fully nonlinear effect of potential features
in the curvature power spectrum.

We begin in \S \ref{sec:curvex} by developing the  Green function approach directly for
curvature excitations which unlike the inflaton excitations can be straightforwardly iterated to 
arbitrary order.   The form of the resultant series is highly constrained by exact relations between
the \bogo excitation coefficients (e.g.~\cite{Greene:2005aj}).
The first true nonlinearity of excitations generating further
excitations occurs only beyond second order  for high frequency features.   For excitations generated at the same epoch, the series can be resummed to give the
fully nonlinear effect of excitations.
In \S \ref{sec:examples} we apply these model-independent techniques to the step and monodromy potentials
and demonstrate that they give accurate predictions for larger than order unity features unlike those in the literature.
We discuss these results in \S \ref{sec:discussion}.

\section{Iterating Curvature Excitations}%%%%%%%%%%%%%%%%%%%%%%%%%%%%%%%%%%%%%%%%%
\label{sec:curvex}

We develop a systematic expansion of comoving curvature fluctuations generated by 
features in the inflaton potential in \S \ref{sec:mode}.  We then relate this expansion to the excitation or
\bogo coefficients of curvature modefunctions to both isolate the superhorizon scale behavior of the curvature fluctuations 
and to exploit the nonlinear relationship between the positive and negative frequency components in
\S \ref{sec:bogo}.   In \S \ref{sec:curvature}, we express the observable curvature power spectrum in terms of
these coefficients  and compare it to second order perturbation theory   for the inflaton
modefunction.   Finally in \S \ref{sec:sub}, we show that for high frequency features of arbitrary amplitude,
the \bogo relation constrains the form of power spectrum features and motivates a nonlinear resummation of the series 
expansion.   This resummation is an exponentiation of the first order excitation  that is exact if the excitations are all generated
at the same epoch and more generally provides a physically motivated template whose parameters may be adjusted for better accuracy.

\subsection{Curvature Modefunctions}%%%%%%%%%%%%%%%%%%%%
\label{sec:mode}

For a canonical scalar inflaton $\phi$, comoving curvature fluctuations $\curv$ obey the Mukhanov-Sasaki equation of motion 
\begin{equation}
\left(\frac{d^2 }{d x^2} - \frac{2}{x} \frac{d }{d x} + 1\right)\curv = -\frac{2}{x}\frac{f'}{f} \frac{d \curv}{d x},
\label{eqn:Reqn}
\end{equation}
where $x = k\eta$ with $\eta=\int_t^{t_{\rm end}} dt/a(t)$ as the conformal time to the end of inflation, 
$'=d/d\ln\eta$ and 
\begin{equation}
f^2 = 4\pi^2 \left( \frac{\dot \phi a \eta}{H} \right)^2 = 8\pi^2 \frac{\epsilon_H}{H^2} ( a H \eta)^2
\end{equation}
as the source of curvature fluctuations.  Here $\epsilon_H$ is the Hubble slow roll parameter.  Curvature fluctuations are related to inflaton field fluctuations in spatially
flat gauge as $\curv = x y/f$, where $y$ satisfies
\begin{equation}
\frac{d^2 y }{ dx^2} + \left( 1 - \frac{2}{ x^2} \right) y = \left( \frac{f''- 3f'}{f} \right)
\frac{y}{x^2} .
\label{eqn:yeqn}
\end{equation}
While Eq.~(\ref{eqn:Reqn}) and (\ref{eqn:yeqn}) are mathematically the same, solving this system in $\curv$ vs.\ $y$ has both practical advantages and disadvantages.
The advantage of using  $y$ is that for a sufficiently large $x$, or equivalently long time before
horizon crossing, the source on the right hand side (rhs)
of Eq.~(\ref{eqn:yeqn}) can be ignored.  
Solutions then correspond to de Sitter modefunctions and the choice of the Bunch-Davies vacuum picks out
and normalizes 
the positive frequency solution
\begin{equation}
\lim_{x\rightarrow \infty} y(x)=y_0(x) \equiv (1+i/x) e^{ix}.
\end{equation}
If variations in the source $f'/f$ become comparable to or faster than the modefunction
oscillation, i.e.~$\Delta \ln \eta \lesssim 1/x$ for $x \gtrsim 1$,  they cause non-adiabatic excitations
of the modefunction out of the vacuum state.
The same simplification of a universal initial form for the modefunction is not true for the curvature since $\curv= xy/f$ depends explicitly on the source $f$.

For this reason,
the original formulation of the generalized slow roll approach  solved Eq.~(\ref{eqn:yeqn})
for the field excitation in $y$  and related these solutions to the curvature fluctuations after horizon crossing \cite{Stewart:2001cd,Choe2004,Dvorkin:2009ne}.   Starting from the universal $y_0$, the GSR approach  replaces
$y$ in the rhs source of Eq.~(\ref{eqn:yeqn}) which then can be solved using Green function techniques.
This procedure is then iterated to improve the solution to the desired order in $f'/f$.
The drawback of this approach is that if the source $f$ continues to vary outside the horizon then so will
the inflaton modefunction since both the lhs and rhs  of Eq.~(\ref{eqn:yeqn}) scale as $y/x^2$.  For potentials
with high temporal frequency sources, this fact would  lead to an apparent breakdown in perturbativity.

On the other hand, the curvature modefunction equation (\ref{eqn:Reqn})
has a superhorizon solution of  $\curv=$\,\,const.~for 
any source.    Ref.~\cite{Dvorkin:2009ne} exploited this fact
by including additional corrections from $f'/f$ at each order in $y$ to keep $\curv = xy/f$ constant for $x\ll 1$.  
This procedure is valid in perturbation theory since these terms already exist at the next order.  However
it is implemented by hand at each order and obscures the reason why the expansion should work when the
inflation modefunction deviates substantially from the de Sitter $y_0$.

To eliminate this superhorizon problem and provide a technique where the iterative improvement is
straightforwardly implemented,  we establish here the equivalent GSR formalism for curvature modefunctions.
This trades the superhorizon problem with an equivalent subhorizon problem  that 
can more easily be finessed.  As discussed above, on subhorizon scales the
curvature depends explicitly on $f$, reflecting the fact that the comoving curvature fluctuation is not defined for a pure de Sitter background.
To finesse this issue, let us exploit the fact that although $f$ can have short timescale transient evolution,
it cannot have a large net evolution across many efolds without ending inflation.   Thus let us assume
that from some initial time $x=x_0$ through horizon crossing $x\ll 1$, $f$ is perturbatively close to some fiducial
value $\fs$.    At the end of the calculation, we cast the final results in a form that is independent of $\fs$ and send $x_0 \rightarrow \infty$.

We can now iteratively improve the solution.
Substituting $\curv \rightarrow \curv_0 \equiv xy_0/f_*$ on the rhs
of Eq.~(\ref{eqn:Reqn}) yields the correction $\curv_1$ and we can repeat this
procedure to find $\curv =\sum_{n=0}^{n_{\rm max}} \curv_n$ to any desired order.   From the Green function, the result is
\begin{eqnarray}
\curv_n(x)& =&\curv_n(x_0)\frac{\curv_0(x)}{\curv_0(x_0)} -i  \int_x^{x_0} \frac{du}{u} \frac{f'}{f} \frac{x}{u} \frac{d\curv_{n-1}}{du}  \nonumber\\
&& 
\left[ y_0^*(u) y_0(x)- y_0^*(x) y_0(u) \right],
\label{eqn:Riterate}
\end{eqnarray}
where the initial value for the homogeneous solution  is
determined by the requirement that  the inflaton modefunction is initially in the Bunch-Davies state,
\begin{eqnarray}
\curv(x_0) &=& \curv_0(x_0) \frac{\fs}{f_0} 
% \nonumber\\ &=&  
= \curv_0(x_0)\sum_{n=0}^\infty \frac{\ln^n (\fs/f_0)}{n!} \nonumber\\
&\equiv& \sum_{n=0}^\infty \curv_n(x_0).
\end{eqnarray}
Here $f_0 \equiv f(\ln x_0)$ and
while we could choose $\fs=f_0$ to eliminate these differences it is useful to keep these quantities 
distinct.  For temporal sources with high frequency
features, we can define $\fs$ as the mean over a few efolds 
rather than the instantaneous value of $f$ since the mean evolves slowly (see \S \ref{sec:sub}).

\subsection{\bogo Coefficients}%%%%%%%%%%%%%%%%%%%%
\label{sec:bogo}
 
It is useful to further characterize the positive and negative frequency components of the excitation
through \bogo coefficients
\begin{eqnarray}
\curv_n(x) = \alpha_n(x) \curv_0(x) + \beta_n(x) \curv_0^*(x).
\end{eqnarray}
Inserting this form into Eq.~(\ref{eqn:Riterate}), we obtain the \bogo hierarchy equations
\begin{eqnarray}
\alpha_n(x) &=& \frac{\ln^n (\fs / f_0)}{n!} \nonumber\\&&
+  \int_x^{x_0}  \frac{du}{u} \frac{f'}{f} y_0^* [ e^{i u}  \alpha_{n-1} - e^{-i u}  \beta_{n-1}], \nonumber\\
\beta_n(x) &=& - \int_x^{x_0}  \frac{du}{u} \frac{f'}{f} y_0  [ e^{i u}  \alpha_{n-1} - e^{-i u}  \beta_{n-1}],
\label{eqn:alphabetaseries}
\end{eqnarray}
with $\alpha_0=1$, $\beta_0=0$. 
Note that we use the fact 
that 
\begin{equation}
\frac{d \curv_n(x)}{d x}  = \alpha_n(x)\frac{d\curv_0(x)}{dx}+ 
\beta_n(x) \frac{d\curv_0^*(x)}{dx}
\label{eqn:Rderiv}
\end{equation}
by virtue of the
Green function construction of Eq.~(\ref{eqn:Riterate}).
To evaluate this series to $n$th order, one simply performs
$n$ one dimensional integrals sequentially as each excitation generates new excitations through their interaction
with the source.

The two \bogo coefficients $\alpha=\sum\alpha_n$, $\beta=\sum\beta_n$ 
are related for any excitation $f'/f$.
This can be seen more directly from the corresponding inflaton modefunction solutions
$y(x)$ and $y^*(x)$.
The Wronskian formed from the two solutions is constant by virtue of Eq.~(\ref{eqn:yeqn}) and given by 
the $x\rightarrow \infty$ Bunch-Davies limit as
\begin{eqnarray}
y \frac{d y^*}{dx} - y^* \frac{dy}{dx} = -2 i.
\end{eqnarray}
Using $y=f \curv/x$ and  Eq.~(\ref{eqn:Rderiv}), we obtain
\begin{equation}
| \alpha(x)|^2 = |  \beta(x) |^2 +  \left( \frac{\fs}{f} \right)^2 ,
\label{eqn:bogrel}
\end{equation}
which differs from the usual \bogo relationship between inflaton modefunction coefficients 
if $f$ evolves away from $f_*$.
This relationship will play a central role in defining the form of the power spectrum for subhorizon 
excitations below.

Finally in order to extract the superhorizon behavior of these excitations, it is useful to form a specific linear
combination of the \bogo coefficients.   Note that 
\begin{eqnarray}
\lim_{x\rightarrow 0} \Re(y_0)& =& -\frac{x^2}{3},  \nonumber\\
\lim_{x\rightarrow 0} \Im(y_0)& =&  \frac{1}{x}.
\end{eqnarray}
By defining $\alpha^\pm = (\alpha \pm \beta)/2$, we isolate the coefficients that multiply
these terms.   Eq.~(\ref{eqn:alphabetaseries}) implies that they obey
\begin{eqnarray}
\alpha^\pm_n(x) 
&=&\frac{1}{2}\frac{\ln^n (\fs/f_0)}{n!}+  \int_x^{x_0} \frac{du}{u} \frac{f'}{f}  [y_0^*(u)\mp y_0(u)] 
\nonumber\\
&&\times \left[  \cos u  \, \alpham_{n-1}+  i \sin u \, \alphap_{n-1}\right].
\label{eqn:alpham}
\end{eqnarray}

By inspection, these coefficients then take the limiting forms
\begin{eqnarray}
\lim_{x\rightarrow 0}\alphap_n &=&\frac{1}{x} {\cal O}  \left( \frac{f'}{f} \right)^n, \nonumber\\
\lim_{x\rightarrow 0}\alpham_n &=& {\cal O}  \left( \frac{f'}{f} \right)^n .
\end{eqnarray}
While $\alphap_n$ diverges outside the horizon, it multiplies the strongly convergent real part of $y_0$ and
hence the superhorizon curvature,
\begin{eqnarray}
\lim_{x\rightarrow 0}  \curv_n 
=  \frac{2 i}{\fs} \alpham_n ,
\label{eqn:Rsuper}
\end{eqnarray}
depends only on $\alpham(0)$.   Likewise though both $\alpha$ and $\beta$ diverge as $1/x$ outside the horizon, their individual effects on the curvature cancel as is consistent with the \bogo relation (\ref{eqn:bogrel}).
A similar issue for the inflaton  \bogo coefficients is even more severe with divergences going
as $1/x^3$ which cancel amongst terms~\cite{Stewart:2001cd,Dvorkin:2009ne}.  This simple 
means of calculating the observable curvature fluctuations provides  an advantage for the curvature coefficient approach over previous iterative approaches.

\subsection{Curvature Power Spectrum}%%%%%%%%%%%%%%%%%%%%
\label{sec:curvature}

The curvature power spectrum simply follows from the superhorizon form of the curvature modefunction in
Eq.~(\ref{eqn:Rsuper})
\begin{equation}
\Delta^2_\curv = \lim_{x\rightarrow 0} |\curv(x)|^2 =
 \frac{\left| 2 \alpham\right|^2}{\fs^2} =
 \frac{1}{\fs^2} \left| 1+  2\sum_{n=1}^{\infty} \alpham_n  \right|^2.
\label{eqn:curviterative}
\end{equation}
For comparison to results from the inflaton modefunction expansion in the literature and their compatibility with the
\bogo relation (\ref{eqn:bogrel}), let us explicitly evaluate the power spectrum
to second order
\begin{eqnarray}
\label{eqn:secondorder}
\ln \Delta_\curv^2 
&=&  -2\ln \fs + 4 \Re(\alpham_1) 
\\
&&\quad + 4[ \Im^2(\alpham_1) - \Re^2(\alpham_1) + \Re(\alpham_2)]+\ldots \nonumber
\end{eqnarray}
The use of the log power spectrum facilitates  comparisons to the literature 
and guarantees a positive definite power spectrum even if the excitations become nonperturbative.

From Eq.~(\ref{eqn:alpham}), the first line gives the zeroth and first order contributions as 
\begin{eqnarray}
\label{eqn:firstorder}
\ln \Delta_\curv^{2(1)}  &=& -2\ln \fs + 2(\ln \fs - \ln f_0)
\nonumber\\
&&\quad  + 4\int_x^{x_0} \frac{du}{u} \frac{f'}{f} \left( \cos^2 u -\frac{\sin 2u}{2u}\right)  \nonumber\\
&=& -2\ln f_0 + 2\int_x^{x_0} \frac{du}{u} \frac{f'}{f} \left[ 1 - W_f(u) \right]  \nonumber \\
&=& -2\ln f - 2\int_x^{x_0} \frac{du}{u} \frac{f'}{f} W_f(u)   , 
\end{eqnarray}
where
\begin{equation}
W_f(u) = \frac{\sin 2u}{u}- \cos 2 u
\end{equation}
with $W_f(0)=1$.

Thus the end result is independent of both the arbitrary fiducial normalization scale $\fs$ and the initial 
conditions $f_0$.  We can therefore take $x_0 \rightarrow \infty$ without loss of generality.
The result also does not depend on the arbitrary superhorizon end point $x \ll 1$ as the integral
simply represents the freezeout of $-2\ln f$ through the window $W_f$.  For example,
for a slowly varying $f=\bar f$, we can interpret this integral as representing a refinement of the leading order
slow-roll freezeout condition $\Delta_\curv^2  = \bar\Delta_\curv^2 \approx 1/\bar f^2|_{x\approx 1}$.  Note that
\begin{equation}
\int_0^{x_0} \frac{du}{u}   \left[ 1 - W_f(u) \right]   \approx \ln x_0 -\ln x_f ,
\label{eqn:ffreeze}
\end{equation}
where 
$\ln x_f= 2 -\gamma_E - \ln 2  \approx 0.7296$ and so for a constant $\ln \bar f'$
\begin{eqnarray}
\ln \bar \Delta_\curv^{2(1)} & \approx &-2 \ln \bar f_0  - 2 (\ln \bar f)' (\ln x_f-\ln x_0) \nonumber\\
&\approx &
 -2 \ln \bar f(\ln x_f).
\label{eqn:Dfreeze}
\end{eqnarray}

Now let us compare the first order result  as derived from inflaton modefunction iteration
\cite{Stewart:2001cd}
\begin{equation}
\ln \Delta_\curv^{2} \approx  G(\ln x) + \frac{2}{3} \int_x^\infty \frac{du}{u} W(u) 
\left( \frac{f''- 3f'}{f} \right) ,
\end{equation}
where 
\begin{eqnarray}
G(\ln x) &=&  -2\ln f + \frac{2}{3}(\ln f)'.
\end{eqnarray}
and 
\begin{equation}
W(u) = {3 \sin 2 u \over 2 u^3} - {3 \cos 2 u \over u^2} - {3 \sin 2 u \over 2 u} .
\end{equation}
Note that $W(0) = 1$, $W(\infty) =0$ and
\begin{equation}
W_f(u) =  W(u)+ \frac{W'(u)}{3}  .
\label{eqn:WfW}
\end{equation}

As noted in Ref.~\cite{Dvorkin:2009ne}, this form is not well defined in that it depends on the arbitrary
evaluation point $x\ll 1$ since the source in the integrand is not the derivative of the boundary term
$G$.   Even if $f'/f$\,=\,const.~as in slow roll this would lead to an unphysical logarithmic 
evolution of the power spectrum outside the horizon.   For a model with high frequency temporal features,
the problem is much worse and can appear as a breakdown in the perturbation expansion near 
horizon crossing.
Ref.~\cite{Dvorkin:2009ne} corrected this problem by replacing 
\begin{equation}
\frac{2}{3}\left( \frac{f''- 3f'}{f} \right) \rightarrow G' = -2(\ln f)' + \frac{2}{3} (\ln f)'',
\label{eqn:replace}
\end{equation} 
which differ by a second order $(f'/f)^2$ term.  
This brings the power spectrum to \cite{Stewart:2001cd}
\begin{eqnarray}
\ln \Delta_\curv^{2({\rm GSR})} \approx G(\ln x) + \int_x^\infty \frac{du}{u} G'(\ln u)  W(u).
\label{eqn:GSR1old}
\end{eqnarray}
As we shall see this term does indeed appear in the second order
expression, and so in perturbation theory, this procedure just amounts to regrouping existing terms.   
However this regrouping lacks an algorithmic formulation for a given term in the iterative series 
and obscures what the criteria is for breakdown of the perturbative expansion.

The curvature modefunction expansion provides a better derivation and justification for this procedure.
Using 
Eq.~(\ref{eqn:WfW}) to 
integrate Eq.~(\ref{eqn:firstorder}) by parts and assuming $x\ll 1$ and $x_0\gg 1$
\begin{eqnarray}
\ln \Delta_\curv^{2(1)} &=&  -2\ln f + \frac{2}{3}(\ln f)' \nonumber\\
&&+ \int_x^{x_0}\frac{du}{u} \left[ -2(\ln f)' +\frac{2}{3} (\ln f)'' \right] W(u) 
\nonumber\\
&=& G(\ln x) + \int_x^{x_0} \frac{du}{u}  G'(\ln u) W(u),
\label{eqn:GSR1}
\end{eqnarray}
which agrees exactly with Eq.~(\ref{eqn:GSR1old})  once $x_0\rightarrow \infty$.

Thus the criteria for perturbative validity is that the curvature modefunction remains close to $\curv\approx \curv_0$,
which is guaranteed outside the horizon if it is close at horizon crossing, as opposed to the
inflaton modefunction  $y\approx y_0$ for all time after horizon crossing.    
Our  new curvature modefunction expansion is explicitly valid out to order
unity curvature power spectrum features and can be straightforwardly iterated to arbitrary order. 

%The old inflaton modefunction approach must be manually corrected
%at each order to gain the same accuracy and cannot be straightforwardly iterated for rapidly varying sources.
%Conversely although sending $x_0\rightarrow \infty$ may seem ill posed from the standpoint of 
%evolution in $f$ causing nonperturbative changes in $\curv$,
 %the validity of this replacement is guaranteed by the fact that inflaton modefunctions 
%are close to de Sitter in this limit.  We also use this fact in \S \ref{sec:sub} to resum the series for slow
%roll evolution in $f$.   

Since the two forms in Eqs.~(\ref{eqn:firstorder}) and (\ref{eqn:GSR1})  are mathematically identical, it is just a matter of convenience
as to which representation to use.   The $G'$ form has some practical advantages since the kernel $W$ decays away
inside the horizon whereas the analogous kernel for $(\ln f)'$ oscillates out to infinity.
Note that under the slow roll assumption of $f'/f \approx$\,const., $G=\bar G$ freezes out at a different epoch than $\bar f$ in determining the power spectrum 
\begin{eqnarray}
\bar\Delta_\curv^{2(1)}  \approx  e^{\bar G(\ln x_G)},
\end{eqnarray}
where $\ln x_G$ can be derived either directly or by matching the alternate form $ e^{-2\ln \bar f(\ln x_f)}$
so that $\ln x_G = \ln x_f+1/3$ (e.g. \cite{Motohashi:2015hpa}).

We can likewise explicitly represent  the second order terms in Eq.~(\ref{eqn:secondorder}).
The first new term is the square of a first order quantity
\begin{eqnarray}
4\Im(\alpham_1) &=&4  \int_x^{x_0} \frac{du}{u} \frac{f'}{f} \left( \cos u\sin u - \frac{\sin^2 u}{u} \right) \nonumber\\
&=& 2 \int_x^{x_0} \frac{du}{u} \frac{f'}{f} \left[ -X(u) - \frac{1}{3} X'(u) \right] \nonumber\\
&\approx &  \int_x^{\infty} \frac{du}{u} G'(\ln u) X(u) \equiv \sqrt{2}  I_1(x),
\end{eqnarray}
where 
\begin{equation}
X(x) = {3 \over x^3} (\sin x- x \cos x)^2 ,
\end{equation}
with $X(0) =  X(\infty) =0$.
The intrinsically second order piece can be isolated as
\begin{eqnarray}
4\Re(\alpham_2) =4 |\alpham_1|^2 + I_2,
\end{eqnarray}
where 
\begin{equation}
I_2 = -4 \int_x^{x_0}  \frac{du}{u} \frac{f'}{f} \left[ X(u) + \frac{1}{3} X'(u) \right] \int_u^{x_0} \frac{dv}{v^2} \frac{f'}{f}.
\label{eqn:I2}
\end{equation}
To second order we obtain 
\begin{equation}
\ln \Delta_\curv^{2(2)}  = G(\ln x) + \int_x^{x_0} \frac{du}{u}  G'(\ln u) W(u) + I_1^2 + I_2,
\label{eqn:secondordergsr}
\end{equation}
which is exactly the same result as derived from the 
curvature conserving replacement procedure of
Eq.~(\ref{eqn:replace}) \cite{Dvorkin:2009ne}.

The nested integral in $I_2$ contains an extra $1/v$ that suppresses contributions from sources
inside the horizon.   In fact had we replaced $y_0 \rightarrow e^{i x}$ in the evaluation of  $\Re(\alpham_2)$,
the $I_2$ term would vanish identically, leaving the whole second order term a sum of squares of first order
terms.   For high frequency transient sources in $f'/f$, these contributions are further suppressed
by integrating to zero whereas modulation by $W(u)$ or $X(u)$ in the first order terms can generate
large excitations out of transient sources as we discuss in the next section.  
Thus as  noted in Ref.~\cite{Dvorkin:2009ne}, for excitations that occur well inside the horizon it is generally a good approximation
to neglect the intrinsically second order term $I_2$.

\subsection{Subhorizon Excitations}%%%%%%%%%%%%%%%%%%%%
\label{sec:sub}

For curvature excitations that are imprinted well before horizon crossing at $x\gg 1$, the general treatment of the
previous sections simplifies considerably providing useful insights about the form of nonlinear excitations in
the power spectrum.

In the \bogo hierarchy equations (\ref{eqn:alphabetaseries}), we can replace $y_0(x) \rightarrow e^{i x}$ in this $x\gg 1$ limit.   The terms that couple  $\alpha_n$ to 
$\alpha_{n-1}$ and $\beta_n$ to $\beta_{n-1}$ involve only the source $f'/f$ whereas those that couple
$\alpha_n$ to $\beta_{n-1}$ and $\beta_n$ to $\alpha_{n-1}$  modulate the source as $e^{\pm 2i x} f'/f$.
Let us first consider why the unmodulated pieces appear.   Suppose we started the modes at $x_0$
with $\alpha_0=1$ and an impulsive excitation that provides a constant  $\beta_1^{({\rm un})}= (f_*/f_0)\beta_{1*}$ where
we have scaled the constant $\beta_{1*}$ for reasons that will be clear below.   With only the
unmodulated terms in Eq.~(\ref{eqn:alphabetaseries}), these would evolve as
\begin{eqnarray}
\alpha_n^{({\rm un})} &=& \frac{\ln^n (\fs / f_0)}{n!} +  \int_x^{x_0}  \frac{du}{u} (\ln f)'\alpha_{n-1}^{(\rm un)} 
=\frac{\ln^n (\fs / f)}{n!}  ,\nonumber\\
\beta_n^{({\rm un})} &=&  \int_x^{x_0}  \frac{du}{u} (\ln f)'\beta_{n-1}^{(\rm un)}
=\frac{\ln^{n-1} (f_0 / f)}{(n-1)!}\frac{\fs}{f_0}\beta_{1*} ,
\end{eqnarray}
and hence resum to 
\begin{eqnarray}
\alpha^{({\rm un})} &=& \sum_{n=0}^\infty  \alpha_n^{({\rm un})} = \frac{ f_*}{f }\alpha_0 ,\nonumber\\
\beta^{({\rm un})} &=& \sum_{n=1}^\infty \beta_n^{({\rm un})} = \frac{ f_*}{f } \beta_{1*}.
\label{eqn:srrenorm}
\end{eqnarray}
Thus we see that these terms in the hierarchy simply renormalize the \bogo coefficients for the net 
evolution in $f$ in accordance with the \bogo relation Eq.~(\ref{eqn:bogrel}).  This evolution does not represent 
a subsequent excitation generated by the original excitation.   For example here $\alpha^{({\rm un})}=f_*/f$
still represents modes in the Bunch-Davies vacuum.
Since the net change in $f$ is responsible for the slow-roll evolution of the curvature modefunctions, 
we will term this nonlinear rescaling ``slow roll renormalization."

The modulated terms in Eq.~(\ref{eqn:alphabetaseries}) represent
excitations generating further excitations.   For rapid variation in $f'/f$, the slow roll renormalization
terms can be small, leading to small overall deviations from scale invariance in the power spectrum 
while the modulated or excitation terms can be large and potentially nonlinear.
While we cannot in general resum this series in closed form,
we can still exploit the \bogo relation (\ref{eqn:bogrel}) to constrain the form of their resummation.
Removing the slow-roll renormalization, we can express the remaining piece in a manner
that makes the \bogo relation (\ref{eqn:bogrel}) manifest
\begin{eqnarray}
|\alpha | &=& \frac{\fs}{f} \cosh\frac{B}{2}, \nonumber\\
|\beta | &=& \frac{\fs}{f} \sinh\frac{B}{2},
\label{eqn:Bbeta}
\end{eqnarray}
leaving an unspecified relative phase between them 
\begin{equation}
\frac{\beta}{\alpha} = e^{i \varphi} \tanh\frac{B}{2}.
\label{eqn:Bbeta2}
\end{equation}

Now let us further assume that all of the excitations occur before horizon crossing so that apart from the
slow roll renormalization,  the source $f'/f$ ceases to change $\alpha$ and $\beta$
or $\alpha^\pm$ 
at some $x \gg 1$.
The power spectrum then takes the form
\begin{eqnarray}
\Delta_{\cal R}^2 &=& \frac{1}{\fs^2} (|\alpha|^2 + |\beta|^2 - 2|\alpha||\beta| \cos\varphi) \nonumber\\
&=&\frac{1}{f^2}  (\cosh B - \sinh B \cos\varphi ).
\label{eqn:gentemplate}
\end{eqnarray}
This form  holds for an arbitrary amplitude excitation $B$ and guarantees a 
positive definite power spectrum since $\cosh B>\sinh B$.
In fact it would continue to hold
for excitations after horizon crossing, but as we have seen in \S \ref{sec:bogo}, $|\beta| \rightarrow \infty$, 
and so $\alpha \rightarrow \beta$ and $\varphi 
\rightarrow 0$.  Here the template form simply cancels to leading order.

Let us now see how this form arises in and illuminates the second order calculation.
To second order,  the \bogo relation (\ref{eqn:bogrel}) gives
\begin{eqnarray}
2 \Re(\alpha_1) & =&  2 (\ln \fs -\ln f), \nonumber\\
2 \Re(\alpha_2) &=& |\beta_1|^2 - |\alpha_1|^2 + 2  (\ln \fs -\ln f)^2 ,%\\
\label{eqn:alphabog}
\end{eqnarray}
which makes the power spectrum
\begin{eqnarray}
\Delta_{\cal R}^{2(2)} &=& \frac{1}{\fs^2} \Big[ 1 + 2\Re( \alpha_1+\alpha_2-\beta_1-\beta_2 -  \alpha_1 \beta_1^*) \nonumber\\
&& \quad +|\alpha_1|^2 + |\beta_1|^2  \Big] \\
&\approx &  \frac{1}{\fs^2} \left[ \left( \frac{\fs}{f} \right)^2 - 2\Re(\beta_1+\beta_2  +  \alpha_1 \beta_1^*)  +2|\beta_1|^2\right]. \nonumber
\end{eqnarray}
In the second line, we have exploited  slow roll renormalization to replace the perturbative
expansion of $f_*/f$ with its nonlinear resummation.   
The only term in the power spectrum that cannot be written  in terms of
first order quantities is $\Re(\beta_2)$.

In the subhorizon excitation limit for $y_0$,
the explicit forms for the coefficients simplify considerably.  For the first order quantities,
\begin{eqnarray}
\alpha_1(x) &=&\ln \fs -\ln f_0 +  \int_x^{x_0}  \frac{du}{u} \frac{f'}{f}   \nonumber\\
&=&
\ln \fs -\ln f  , \nonumber\\
 \beta_1(x) &=& - \int_x^{x_0}  \frac{du}{u} \frac{f'}{f}  e^{2 i u} .
 \label{eqn:alphabeta1}
\end{eqnarray}
Note that $\alpha_1$ is consistent with the \bogo relation for $\Re(\alpha_1)$ in Eq.~(\ref{eqn:alphabog}) and is itself a real quantity.
It appears as the first term in the slow roll renormalization of  $\alpha_0=1$ to $\alpha=f_*/f$.

Using these quantities in Eq.~(\ref{eqn:alphabetaseries}) to define the second order quantities gives
\begin{eqnarray}
\alpha_2(x) &=& \frac{(\ln \fs -\ln f)^2}{2} +  \int_x^{x_0}  \frac{du}{u} \frac{f'}{f}e^{-2i u} \int_u^{x_0} \frac{dv}{v}  \frac{f'}{f}e^{2i v}  ,
\nonumber\\
\beta_2(x) &=& (\ln \fs-\ln f_0) \beta_1 -   \int_x^{x_0}  \frac{du}{u} \frac{f'}{f}e^{2i u} \int_u^{x_0}   \frac{dv}{v}  \frac{f'}{f}
\nonumber\\
&&\quad  - \int_x^{x_0}  \frac{du}{u} \frac{f'}{f}\int_u^{x_0}   \frac{dv}{v}  \frac{f'}{f}e^{2i v}  \nonumber\\
&=& \alpha_1 \beta_1.
\end{eqnarray}
Since 
\begin{align}
2 \Re\left[ \int_x^{{x_0}} du F(u)  \int_u^{{x_0}} {dv}F^*(v) \right]= \left| \int_x^{{x_0}} d u F(u) \right|^2,
\end{align}
we recover Eq.~(\ref{eqn:alphabog}) for $\Re(\alpha_2)$ and in the subhorizon excitation limit $\Re(\beta_2)$ 
can also be written in terms of first order quantities. This is of course just a restating of the fact that the
$I_2$ contribution in Eq.~(\ref{eqn:I2}) is negligible and comes from the deviation of $y_0(u)$ from $e^{i u}$.  In fact $\alpha_1$ appears here
because it is the first term in the slow roll renormalization of $\beta_1$,  $\beta_1 +\alpha_1\beta_1 \approx (f_*/f)\beta_1$ 
and each further order will contain terms that resum to the full correction as shown in Eq.~(\ref{eqn:srrenorm}).

We can combine these terms back into the template form of Eq.~(\ref{eqn:gentemplate}), 
\begin{eqnarray}
\Delta_{\cal R}^{2(2)} &\approx& \frac{1}{\fs^2} \left[\left( \frac{\fs}{f} \right)^2 + 2|\beta_1|^2 - 2\left(\frac{\fs}{f} \right)^2 \Re(\beta_1)\right] \nonumber\\
&\approx &\frac{1}{f^2}[\cosh(2|\beta_1|)  -\sinh (2|\beta_1|)  \cos\varphi],
\label{eqn:Bform1}
\end{eqnarray}
where we have inserted the appropriate cubic and higher order terms that appear due
to slow roll renormalization  to factor out $(f_*/f)^2$.
%
%\vin{In the first line - last term - I didn't get where the power 2 came from. I got $-2 (f_*/f) \mathcal{R}(\beta_1)$. And from first to second line I didn't get how to rearrange $(f_*/f)^2 + 2|\beta_1|^2$ as the taylor expansion of the $cosh$ without the prefactor $(f_*/f)^2$ in front of  $2|\beta_1|^2$}
We can now  associate  $B=2 |\beta_1|$ and $\varphi$ with the absolute phase  of $\beta_1=e^{i\varphi} B/2$.   
Phrased in this form, the first order calculation of the power spectrum is in fact accurate to second order.

Eq.~(\ref{eqn:Bform1}) automatically accounts for the resummed, slow roll renormalization of 
$f$ due to its evolution from the excitation epoch to horizon crossing.  
To make the correspondence between the nonlinearly resummed renormalization and the first order excitation general and explicit, let us model
\begin{equation}
\ln f = \ln \bar f + \delta\ln f,
\end{equation}
where we take a log-linear evolution
\begin{eqnarray}
\ln \bar f
&=& \ln \fs + \frac{n_s-1}{2}[ \ln (k_0 \eta)-\ln x_f] \nonumber\\ 
&=&  \ln \fs + \frac{n_s-1}{2}[\ln x - \ln(k/k_0) -\ln x_f ].
\end{eqnarray}
Here the slope, or as we shall see the power spectrum tilt,  $n_s=$\,\,const., $k_0\eta=x_f$ defines the zero point, 
and the freezeout point $x=x_f$ is given by Eq.~(\ref{eqn:ffreeze}).   Through Eq.~(\ref{eqn:Dfreeze}), the model for $\ln \bar f$  defines a
power spectrum that is  a power law to leading order
\begin{eqnarray}
\bar I_0 &\equiv &  \ln \bar \Delta_\curv^2 
=-2\ln \fs + (n_s-1) \ln (k/k_0),
\end{eqnarray}
which can be used to replace the slow roll term  in Eq.~(\ref{eqn:Bform1}) $1/f^2 \rightarrow e^{ \bar I_0}$. 
For notational convenience let us
define
\begin{eqnarray}
\delta I_0 &=&  \ln \Delta_\curv^{2(1)}  - \ln \bar\Delta_\curv^2 \\
&=& -2\delta \ln f_0
+4 \int_x^{x_0} \frac{du}{u} (\delta \ln f)'   \left(\cos^2 u -\frac{\sin 2u}{2u}\right) ,\nonumber
\end{eqnarray}
and similarly
\begin{eqnarray}
\delta I_1& =&  I_1 - \bar I_1 \\
&=& 2\sqrt{2}  \int_x^{x_0} \frac{du}{u} (\delta \ln f)'   \left( \cos u\sin u - \frac{\sin^2 u}{u} \right), \nonumber
\end{eqnarray}
where 
\begin{equation}
\bar I_1 = \frac{\pi}{2\sqrt{2}}(1-n_s) .
\end{equation}
Comparing with Eq.~(\ref{eqn:alphabeta1}) at $u\gg 1$, we see that $\delta I_0$ and $\delta I_1$ capture the
subhorizon excitation as
\begin{equation}
4\alpham_1= 2(\alpha_1-\beta_1) \rightarrow \delta I_0 +i\sqrt{2}\delta I_1,
\label{eqn:I1I2corr}
\end{equation}
whereas $\bar I_0$ captures the slow-roll renormalization of $\alpha_1$ and $\beta_1$ from the excitation.  
The second order power spectrum for subhorizon excitations then {becomes}
\begin{equation}
\Delta_{\cal R}^{2(2)} = \bar \Delta_{\cal R}^{2}\left[ \bar I_1^2 + \bar I_2 + \cosh B
- \sinh B \cos\varphi \right],
\label{eqn:gen2form}
\end{equation}
where $\bar I_2 = -(n_s-1)^2$ and
\begin{eqnarray}
B^2 &\approx &  \delta I_0^2 + 2 \delta I_1^2, \nonumber\\
\cos \varphi &\approx & - \frac{  \delta I_0\cos\bar\varphi -\sqrt{2} \delta I_1\sin \bar\varphi }{ \sqrt{\delta I_0^2  + 2\delta I_1^2}}\nonumber\\
&\equiv & \cos(\bar\varphi + \delta\varphi),
\label{eqn:Bvarphi}
\end{eqnarray}
with 
\begin{equation}
\bar\varphi =-\sqrt{2}\bar I_1= -\frac{\pi}{2}(1-n_s).
\end{equation}
We include $\bar I_1$ as a phase shift $\bar \varphi$ since it can alternately be derived by keeping track of the phase difference between
the positive and negative frequency components using {Hankel} functions for slow-roll modefunctions in the presence of tilt.  
This ensures that its effect is well modeled even for high amplitude excitations where the second order approximation
has broken down.  Finally in Eq.~(\ref{eqn:gen2form}), the correction terms to the leading order 
slow roll power spectrum nearly cancel
\begin{equation}
\bar I_1^2 + \bar I_2 = \left(\frac{\pi^2}{8} -1\right) (1-n_s)^2,
\end{equation}
and for viable $1-n_s \approx 0.03 $ produce $\sim 10^{-4}$ corrections which we can typically neglect.

To summarize the subhorizon second order results, nonlinear effects from excitations generating excitations only appear  through
$\delta I_2$ which is negligible for subhorizon high-frequency excitations.   
This is a consequence of the \bogo relation (\ref{eqn:bogrel}) for $\Re(\alpha_2)$ and the fact that
$\beta_2$ is simply a rescaling of the linear excitation $\beta_1$.

Thus the first true effect of subhorizon excitations generating further excitations occurs at third order.
Even for subhorizon effects, third order excitations do not reduce exactly to products of first order excitations.  In 
particular since
\begin{eqnarray}
\beta_3(x) &\approx & - \int_x^{x_0}  \frac{du}{u} \frac{f'}{f} [e^{2i u}  \alpha_{2} -  \beta_{2}],
\end{eqnarray}
the cubic excitation involves $\Im(\alpha_2)$, which is not determined by the \bogo relation.  On 
the other hand, these higher order coefficients are repeated integrations of the same fundamental
modulated source of excitation $e^{2iu} f'/f$ that is responsible for $\beta_1$.   
As shown in Eq.~(\ref{eqn:srrenorm}), the
unmodulated $f'/f$ pieces are responsible for slow roll renormalization.
Thus the modulated  or excitation part of the hierarchy is a function of the first order excitation function $\beta_1(u)$
\begin{eqnarray}
\alpha_n(x) &\approx &- \int_x^{x_0}  \frac{du}{u} \beta_1'^*(u) \beta_{n-1}(u), \nonumber\\
\beta_n(x) &\approx&- \int_x^{x_0}  \frac{du}{u} \beta_1'(u) \alpha_{n-1}(u) .
\label{eqn:recurse}
\end{eqnarray}
If we take  $\beta_1= e^{i\varphi} |\beta_1| $ and further assume $\varphi\approx$\,const. as would be
the case if all excitations were generated at one epoch,
\begin{eqnarray}
\alpha_n(x) &\approx &
	\begin{cases} |\beta_1|^n/n!  & \quad (n={\rm even}),\\
	0 & \quad (n={\rm odd}), \\
	\end{cases}	\nonumber\\
\beta_n(x) &\approx &
	\begin{cases}
	0 & \quad (n={\rm even}),\\
	e^{i\varphi} |\beta_1|^n/n!  & \quad (n={\rm odd}) ,\\
	\end{cases}
\label{eqn:resummation}
\end{eqnarray}
which resums into 
\begin{eqnarray}
\alpha &\approx & \cosh |\beta_1|, \nonumber\\
\beta &\approx &  e^{i\varphi} \sinh |\beta_1|.  
\end{eqnarray}	
Thus we expect excitations to generate further excitations leading to an exponentiation of the linearized
effect on the power spectrum.  In this constant phase approximation,  Eq.~(\ref{eqn:Bvarphi}) for  $B$ and $\varphi$ 
associated with the linear calculation are the fully nonlinear relation 
for the template power spectrum
\begin{equation}
\Delta^{2(\NL)}_{\curv} \approx \bar \Delta_{\cal R}^{2}(\cosh B -\sinh B \cos\varphi).
\label{eqn:template}
\end{equation}
Beyond cases
where the constant phase approximation holds, we  call the combination of Eqs.~(\ref{eqn:Bvarphi}) and (\ref{eqn:template}) the nonlinear (NL) ansatz.
Errors induced by this prescription take the form of changes
to $B(k)$ and $\varphi(k)$ rather than extra terms that make the power spectrum unphysical in single field
inflation.

This ansatz should be compared with
a similar exponentiation of $\delta I_0$  proposed in Ref.~\cite{Dvorkin:2009ne}
\begin{eqnarray}
\Delta^{2({\rm GSR1})}_{\curv} &\equiv & e^{I_0} (1+I_1^2 )\nonumber\\
&=& \bar \Delta_\curv^2 e^{\delta I_0} [1 + (\bar I_1+\delta I_1)^2],
\end{eqnarray}
which equally well satisfies the second order form Eq.~(\ref{eqn:secondordergsr})
for subhorizon excitations and preserves a positive definite power spectrum but differs in not explicitly
enforcing the \bogo relation (\ref{eqn:bogrel}).  It is therefore limited in accuracy to order
unity excitations $\sqrt{2} |\delta I_1|<1$ \cite{Dvorkin:2010dn} whereas Eq.~(\ref{eqn:template}) is not.
We shall next compare the NL ansatz to the exact result for explicit examples of potential features.

\section{Featured Examples}%%%%%%%%%%%%%%%%%%%%%%%%%%%%%%%%%%%%%%%
\label{sec:examples}

We apply the formalism developed in the previous section to two examples of high frequency features:
sharp steps (\S\ref{sec:steps}) and axion monodromy oscillations (\S\ref{sec:monodromy}) in the inflaton potential.
In the sharp step limit, all excitations are generated at the same epoch and the nonlinear resummation
of the excitation hierarchy in the template form of Eq.~(\ref{eqn:template}) is exact for subhorizon modes.   
Deviations due to the finite duration of the excitation can be characterized 
by small changes in the parameters of the template form.  
For monodromy, in addition to excitations contemporaneously generating excitations that can be
resummed, excitations resonantly generate excitations over an extended period of time.  
These too may be computed from cubic and higher order terms in the hierarchy and mainly produce changes in
the parameters of the template form.

\subsection{Sharp Steps}%%%%%%%%%%%%%%%%%%%%
\label{sec:steps}

A sharp step in the potential provides a simple example where the excitation hierarchy ($\alpha_n$, $\beta_n$)
of Eq.~(\ref{eqn:alphabetaseries}) and hence Eq.~(\ref{eqn:template}) can be explicitly calculated in closed form.  
We model the step potential as
\begin{equation}
V(\phi) = \bar V(\phi) \{ 1+ 2 b_V [S(\phi-\phi_s)-1]\},
\end{equation}
where $ -\infty < b_V < 1/2$ and $S$ is a step function defined to be 0 before the step at $\phi_s$ and 1 after the step.
% so that the potential goes from $\bar V(1-2b_V)$ to $\bar V$ across the step.  
Note that $b_V>0$ therefore is a step up, which is bounded by stepping from zero, and $b_V<0$ a step down to a fixed $\bar V(\phi) $, 
which is unbounded.   Here $\bar V(\phi)$ is an underlying smooth slow-roll potential.   In numerical comparisons,
we take $\bar V(\phi)= V_0 (1-\beta\phi^2/6)$ for definiteness.

\subsubsection{Nonlinear Excitations}%%%%%%%%%%

In order to evaluate the excitation hierarchy, we need a model for $(\delta \ln f)'$.    Since $f^2 \propto \epsilon_H \propto \dot\phi^2$ with
the time evolution of other terms subdominant, this amounts to understanding the change in the kinetic
energy of the inflaton due to rolling over a step.    
Following Ref.~\cite{Miranda:2013wxa}, we can start by using energy conservation to model the jump across the step
\begin{eqnarray}
\Delta\ln f &\equiv &  \frac{1}{2} \Delta\ln \epsilon_H
 =  \frac{1}{2} (\ln \epsilon_{HI} - \ln \epsilon_{HB} )\nonumber\\
&=&
\frac{1}{2} \ln\left( 1- 6 b_V/\epsilon_{HB} \right),
\end{eqnarray}
where  $\epsilon_H=\epsilon_{HB}$ before the step and $\epsilon_{HI}$ immediately after the step. Given that the step is sharp 
\begin{equation}
(\delta \ln f )'=( \Delta\ln f )S',
\end{equation}
where note that $' =d/d\ln\eta \approx -d/dN$.  
Integrating to larger values in $\ln x$  goes from after the step to before the step which changes $S$  by $-1$.  
Thus $S'$ is the negative of a delta function
\begin{equation}
\int_{x}^\infty \frac{du}{u} S' = - S(x).
\end{equation}

Using this model we can evaluate the $\alpha$ and $\beta$ hierarchy of coefficients for a sharp step.  
We assume that the modes in question encounter the step deep within the horizon $x_s=k \eta(N_s) \gg 1$ and
that the width of the step $\delta x_s = k\delta \eta(N_s) \ll 1$.   
We will return to violations of these assumptions at low and high $k$ below.  
Starting with $\alpha_0=1$, $\beta_0=0$, we can evaluate the first order excitations with Eq.~(\ref{eqn:alphabetaseries}),
\begin{eqnarray}
\alpha_1 &=& (-\Delta \ln f)S, \nonumber\\
\beta_1 &=& -e^{2 i x_s} (-\Delta \ln f)S .
\end{eqnarray}
Since each successive term adds a factor of $S$ which is then multiplied by $S'$ from the source $(\ln f)'$, 
the integrals reduce to evaluating
\begin{equation}
\int_x^\infty \frac{du }{u} S^{n-1} S' =\frac{1}{n} \int_x^\infty \frac{du}{u} \frac{d S^n}{d\ln u} = -  \frac{1}{n} S^n(x).
\end{equation}
Thus the hierarchy of \bogo coefficients is given explicitly by
\begin{eqnarray}
\alpha_n &=& \frac{2^{n-1}}{n!} (-\Delta\ln f S)^n,  \nonumber\\
\beta_n &=& -  \frac{2^{n-1}}{n!} (-\Delta\ln f S)^n e^{2 i x_s}.
\end{eqnarray}
%Note that unlike monodromy, there is no relative phase shift between orders since this excitation is generated entirely at
%one epoch $\eta_s$ in the limit of a sharp step.   For a finite width step we would expect a phase shift for $k$-modes where
%the step is resolved, i.e. at $x_s>x_d$ in the damping tail of the excitation.

Summing the series, we have immediately after the step
\begin{eqnarray}
\alpha_I &\equiv& \sum_n\alpha_n= \frac{1}{2} + \frac{1}{2} e^{-2 \Delta \ln f}, \nonumber\\
\beta_I &\equiv&\sum_n\beta_n= \left(  \frac{1}{2} - \frac{1}{2} e^{-2 \Delta \ln f} \right) e^{2 i x_s} .
\end{eqnarray}
Note that this resummation satisfies the \bogo relation (\ref{eqn:bogrel})
\begin{equation}
|\alpha_I|^2 -|\beta_I |^2 = e^{-2\Delta\ln f} 
=\left( \frac{ f_B}{f_I} \right)^2.
\end{equation}

Now we need to evolve the coefficients from the step through to freezeout.   
The kinetic energy imparted by the step decays back to the attractor value after several efolds.   
On the attractor  $\bar\phi_{,N} \approx -V_{,\phi}/3 H^2$ and so the excess 
$\delta\phi_{,N} = \phi_{,N} - \bar\phi_{,N}$ evolves under the Klein-Gordon equation as
\begin{equation}
\delta \phi_{,NN} = -3 \delta\phi_{,N}. 
\end{equation}
Thus this excess decays with efolds as
\begin{equation}
\delta\phi_{,N} \propto e^{-3(N-N_s)}\end{equation}
where $N_s$ is the efold at which the inflaton encounters a step.   This generalizes the 
treatment of Ref.~\cite{Miranda:2013wxa} to extremely large amplitude steps.

Since $\epsilon_H =(\phi_{,N})^2/2$,  after many efolds $f \rightarrow f_A = f_B$.   
Since this evolution is slow compared with $\Delta x=1$, it simply renormalizes the coefficients as discussed in \S \ref{sec:sub}
\begin{eqnarray}
\alpha_A &=& \frac{f_I}{f_A} \alpha_I 
= e^{\Delta\ln f} \alpha_I 
= \cosh(\Delta\ln f), \nonumber\\
\beta_A &=& \frac{f_I}{f_A} \beta_I = e^{\Delta\ln f} \beta_I 
=\sinh(\Delta\ln f) e^{2 i x_s},
\end{eqnarray}
which again satisfies the \bogo relation
\begin{eqnarray}
|\alpha_A|^2 -|\beta_A |^2 &=& \cosh^2(\Delta\ln f) - \sinh^2(\Delta\ln f) \nonumber\\
&=& 1 = \left( \frac{f_B}{f_A} \right)^2 .
\end{eqnarray}
From Eqs.~(\ref{eqn:Bbeta}) and (\ref{eqn:Bbeta2}), we can read off the nonlinear template amplitude and phase
\begin{eqnarray}
B &=& 2\Delta \ln f =  \ln\left( 1- 6 b_V/\epsilon_{HB} \right), \nonumber\\
\delta\varphi &=& 2 x_s.
\end{eqnarray}
There is no restriction on the amplitude of the step $b_V$, save 
% save is a synonym for except
that if $6 b_V>\epsilon_{HB}$ the
initial kinetic energy is insufficient to carry the inflaton over the step up.

This description suffices for modes that were deep inside the horizon when the inflaton rolled over the
step but not so deep that the mode oscillates during the transition.   Using the first-order-based template prescription for 
$B$ of Eq.~(\ref{eqn:Bvarphi}) we can extend the description into these regimes.
As in Eq.~(\ref{eqn:I1I2corr}), we replace $2(\alpha_1-\beta_1)$ with
\begin{eqnarray}
\delta I_0 &=& \frac{2\Delta\ln f}{3} {\cal D}\left(\frac{x_s}{x_d}\right)W'(x_s) ,\nonumber\\
\sqrt{2}\delta I_1 &=& \frac{2\Delta\ln f}{3} {\cal D}\left(\frac{x_s}{x_d}\right)X'(x_s).
\label{eqn:I0step}
\end{eqnarray}
Following Ref.~\cite{Adshead:2011jq,Miranda:2013wxa} we account for a finite width step with a damping
function ${\cal D}$, which for a step shape
\begin{equation}
S = \frac{1}{2} \left[ \tanh\left(\frac{\phi-\phi_s}{d} \right) + 1\right]
\end{equation}
is given by
\begin{equation}
{\cal D}(y) = \frac{y}{\sinh y}.
\end{equation}
Here the damping scale 
\begin{equation}
x_d = \frac{ \phi'}{\pi d},
\end{equation}
and we have assumed for definiteness that the inflaton rolls over the step at $\phi_s$ toward larger field values.
Thus the amplitude parameter becomes
\begin{eqnarray}
B&=& \sqrt{\delta I_0^2 + 2\delta I_1^2}  \\
&=&  \ln\left( 1- \frac{6 b_V}{\epsilon_{HB}} \right){\cal D}\left(\frac{x}{x_d}\right)\frac{\sqrt{ {W'}^2(x_s)+{X'}^2(x_s) }}{3}
\nonumber
\end{eqnarray}
and the phase
\begin{equation}
\cos\varphi =- \frac{W'(x_s)\cos\bar\varphi - X'(x_s)\sin\bar\varphi}{\sqrt{{W'}^2(x_s) + {X'}^2(x_s)}}.
\end{equation}
This prescription exactly coincides with that given in Ref.~\cite{Miranda:2013wxa} to second order in the perturbation 
to the kinetic energy $6b_V/\epsilon_{HB}$ and generalizes it to arbitrarily large steps by matching it to the
fully nonlinear calculation in the regime $1\ll x_s \ll x_d$.

\begin{figure}[t] %------------------------------
\psfig{file=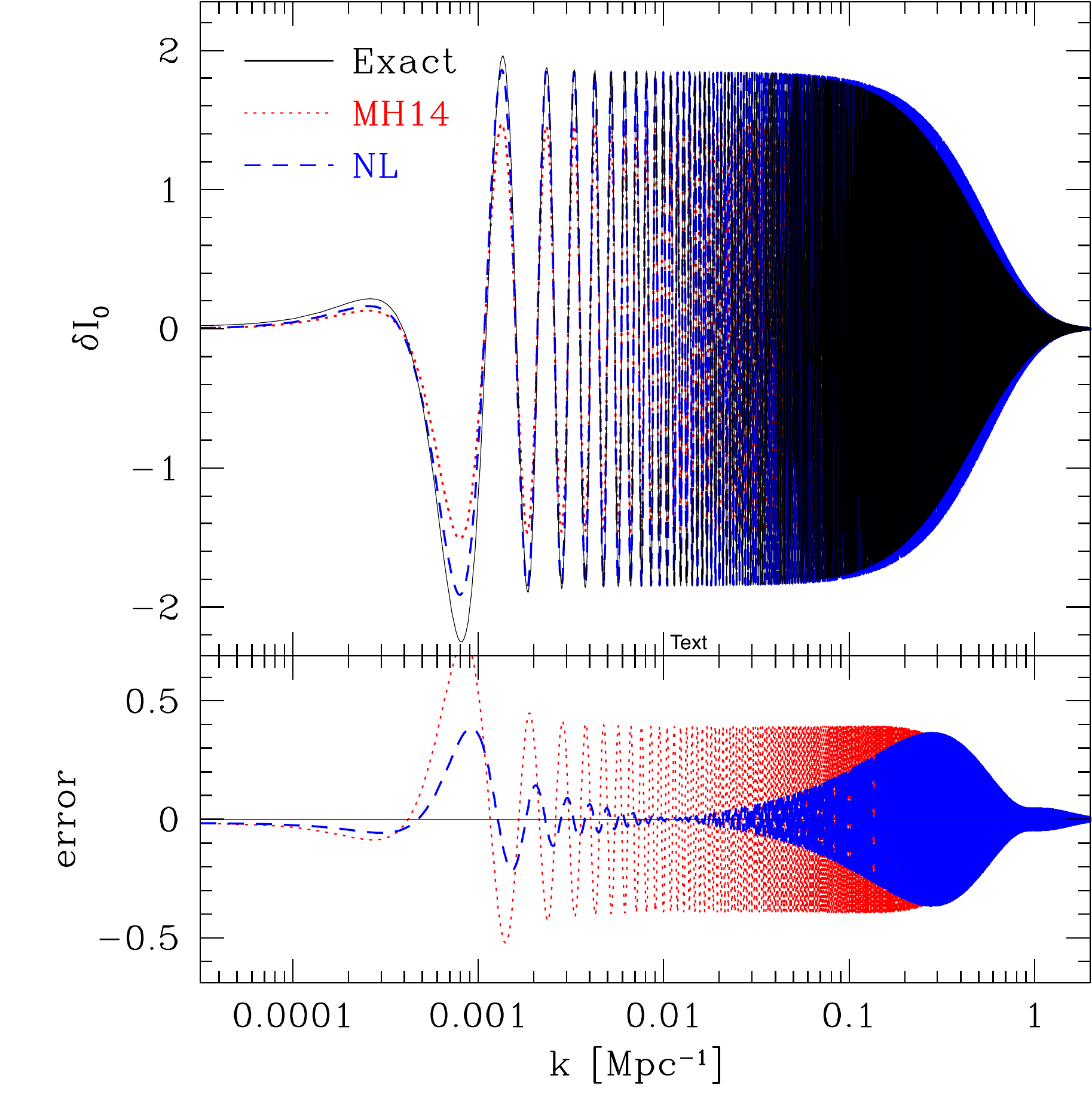, width=3.25in}
\caption{\footnotesize First order curvature response $\delta I_0$ to an extremely large, sharp step:
$6 b_V/\epsilon_{H B} = -5.34$, $x_d \approx  750$ at $\eta_s = 3299.780$ Mpc.   Compared are
the  numerical result of integrating the source $f'/f$ (exact, solid), the analytic second order
approximation from Ref.~\cite{Miranda:2013wxa} (MH14, dotted) and the nonlinear
analytic approximation of this work (NL, dashed) with differences with exact highlighted (bottom panel).   In the region between the
start of the oscillations and their damping, the NL provides a highly accurate form whereas
MH14 is discrepant at order unity.}
 \label{plot:step_I0}	
\end{figure} %------------------------------

\subsubsection{Comparisons and Fits}%%%%%%%%%%

In Fig.~\ref{plot:step_I0} we show an example with a very large step $6 b_V/\epsilon_{H B} = -5.34$  which the 
inflaton crosses when the horizon is comparable to the current horizon $\eta_s = 3299.780$ Mpc.
We first compare $\delta I_0$ as calculated exactly from $f(\ln \eta)$, our nonlinear calculation
of Eq.~(\ref{eqn:I0step}) and the second order calculation of Ref.~\cite{Miranda:2013wxa}.
Although $\delta I_0$ is the first order response of the curvature modefunctions to the source, the source
itself is nonlinear in the sense of imparting a large change in the kinetic energy of the inflaton.
In the region  $1\ll x_s \ll x_d$, our nonlinear (NL) analytic calculation accurately models the amplitude of the oscillations
whereas that of Ref.~\cite{Miranda:2013wxa} (MH14) does not.   Here $x_d \approx  750$ and $x_s=1$ at $k \approx 0.0003$ Mpc$^{-1}$.
We follow the prescription in Ref.~\cite{Miranda:2013wxa} for relating these parameters to those of the potential.

\begin{figure*}[t] %------------------------------ 
\psfig{file=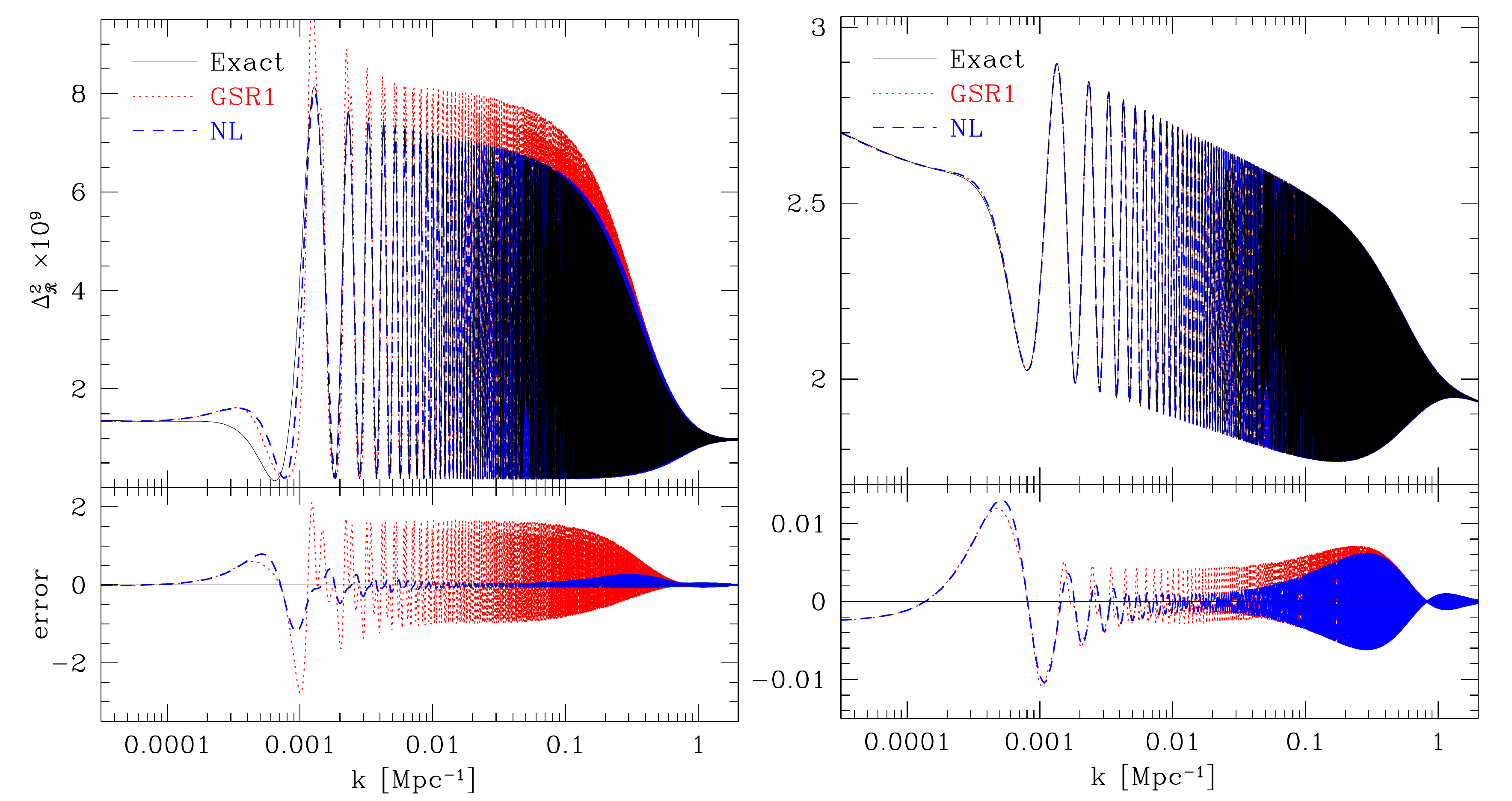, width=6.25in} 
\caption{\footnotesize Curvature power spectrum $\Delta_\curv^2$ for
the extremely large step of Fig.~\ref{plot:step_I0} (left,  $6 b_V/\epsilon_{H B} =-5.34$, 
$\eta_s = 3299.78$ Mpc)
and a large step (right, $6 b_V/\epsilon_{H B}=0.2$, $\eta_s = 3299.79$ Mpc) each with
$x_d  \approx 750$.  Compared are the numerical result (exact, solid), the GSR1 second order
form of Eq.~(\ref{eqn:GSR1}) and the nonlinear form of Eq.~(\ref{eqn:template}).   Both analytic
forms use the same nonlinear analytic calculation of $\delta I_0$ and $\delta I_1$ from
Fig.~\ref{plot:step_I0}.   Even at extremely large amplitudes the NL form provides
a highly accurate description whereas GSR1 using the same first order responses does not.
At amplitudes below unity, both  perform well but NL generally exceeds the accuracy of GSR1.}
\label{plot:step_curv}	
\end{figure*} %------------------------------

Even with the correct nonlinear $\delta I_0$, the GSR1 second order based form of Eq.~(\ref{eqn:GSR1}) 
errs in predicting the power spectrum.   In Fig.~\ref{plot:step_curv}, we compare the exact power spectrum to
the GSR1 form and our new nonlinear form.  The NL again agrees well with the exact form for $1\ll x_s \ll x_d$ 
whereas even with the better analytic calculation of $\delta I_0$ from Fig.~\ref{plot:step_I0}, GSR1 does not.
We also show a smaller step ($6 b_V/\epsilon_{H B} = 0.2$) where the second order GSR1
approximation should be valid.   In this case both GSR1 and NL perform well but NL is
still markedly better in the region  $1\ll x_s \ll x_d$.

\begin{figure*}[t] %------------------------------ 
\psfig{file=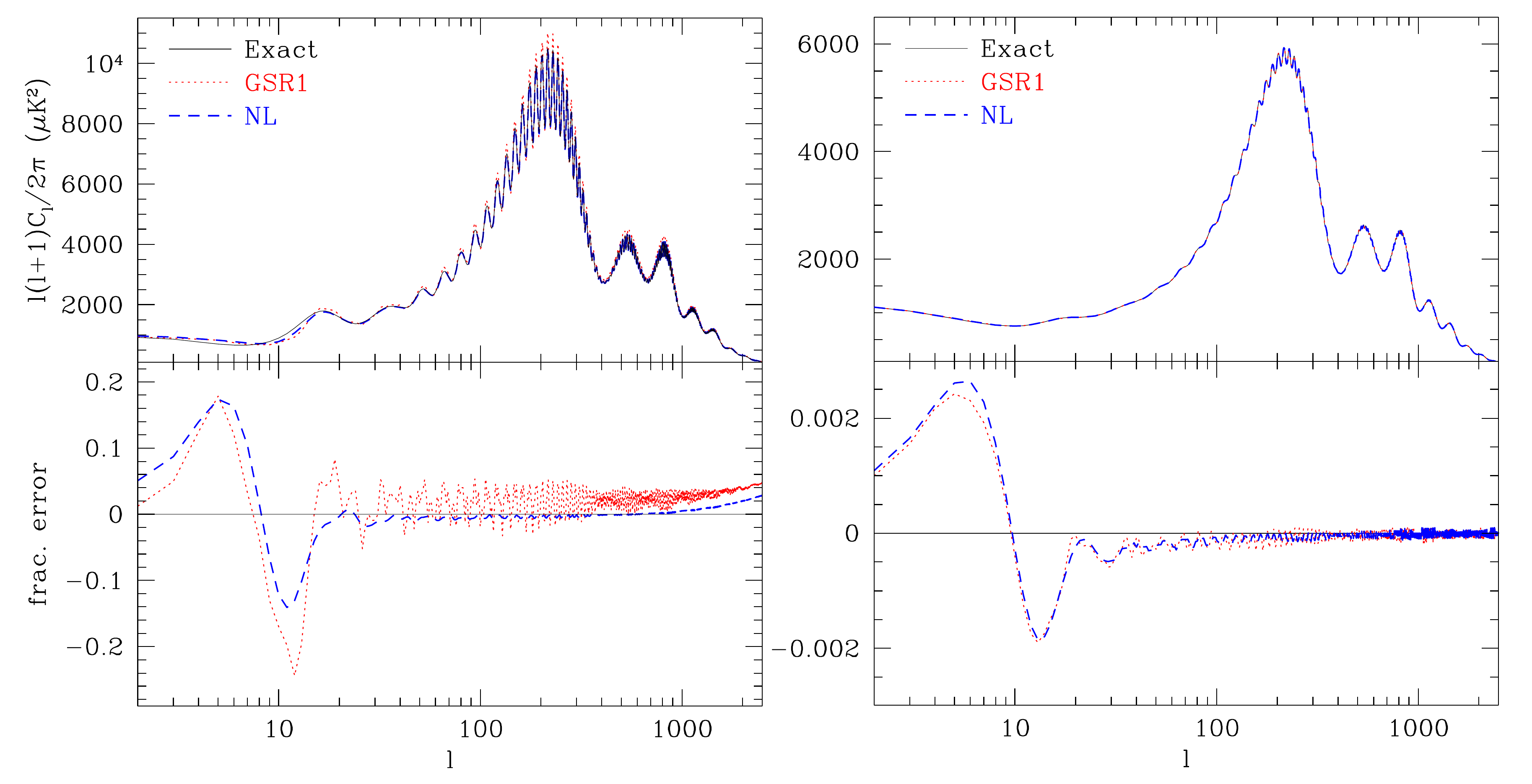, width=6.25in}
\caption{\footnotesize CMB temperature power spectrum $C_\ell$ for the extremely large
step and moderately large step cases of Fig.~\ref{plot:step_curv}.   
Errors in $C_\ell$ are even smaller than in $\Delta_\curv^2$ due to projection effects.}
\label{plot:step_Cl}	
\end{figure*} %------------------------------

The errors in the CMB power spectrum are even smaller due to projection effects as shown in Fig.~\ref{plot:step_Cl}.   
Because projection effects also fill in the power at the oscillation troughs, 
here we plot fractional errors for an easier comparison with cosmic variance limits.  
Even for the large amplitude step the errors for $x\ll x_d$ are comparable
to or smaller than cosmic variance errors through the damping tail. For the smaller amplitude step, both are.

\begin{figure}[t] %------------------------------
\psfig{file=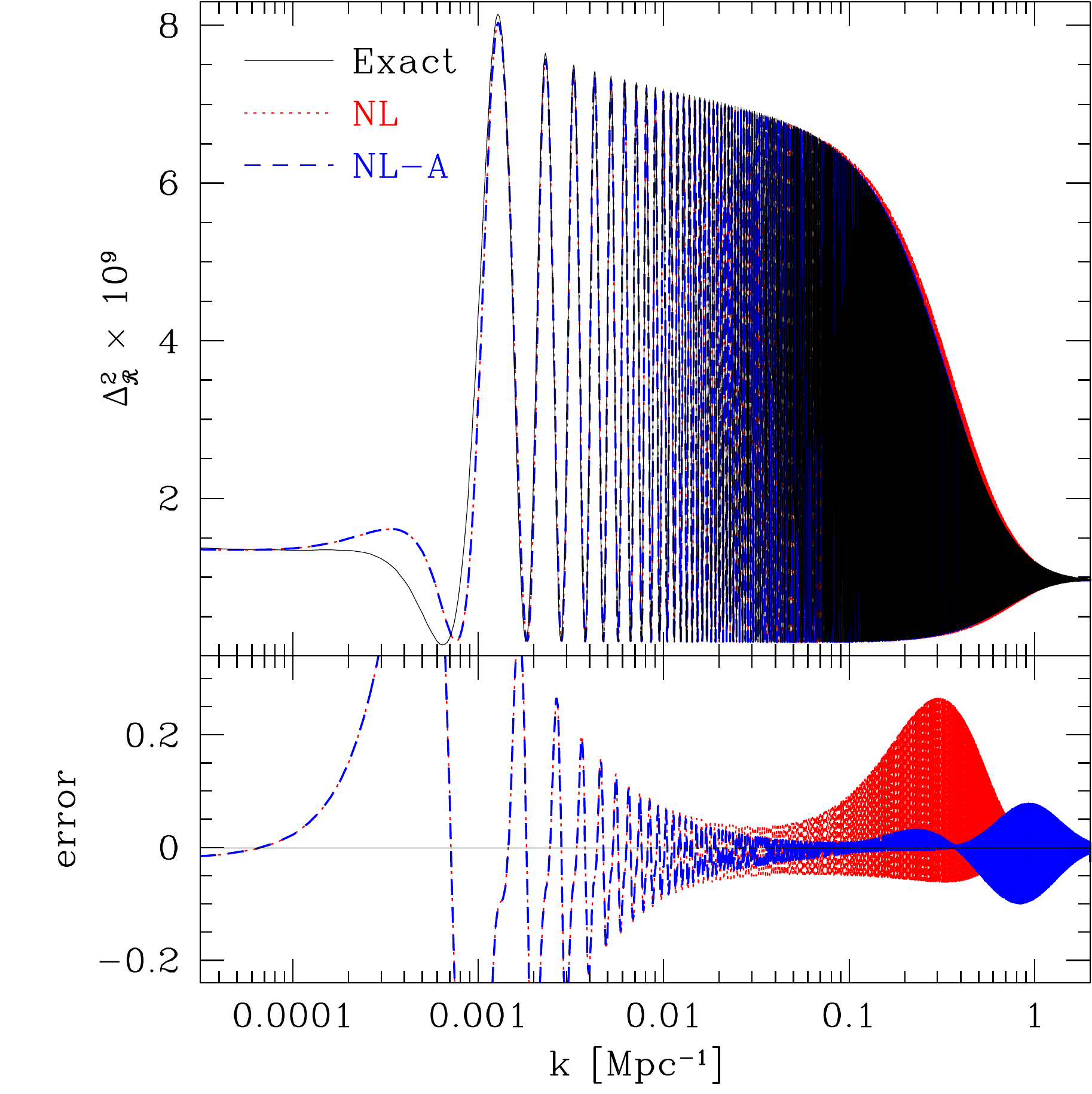, width=3.25in}
\psfig{file=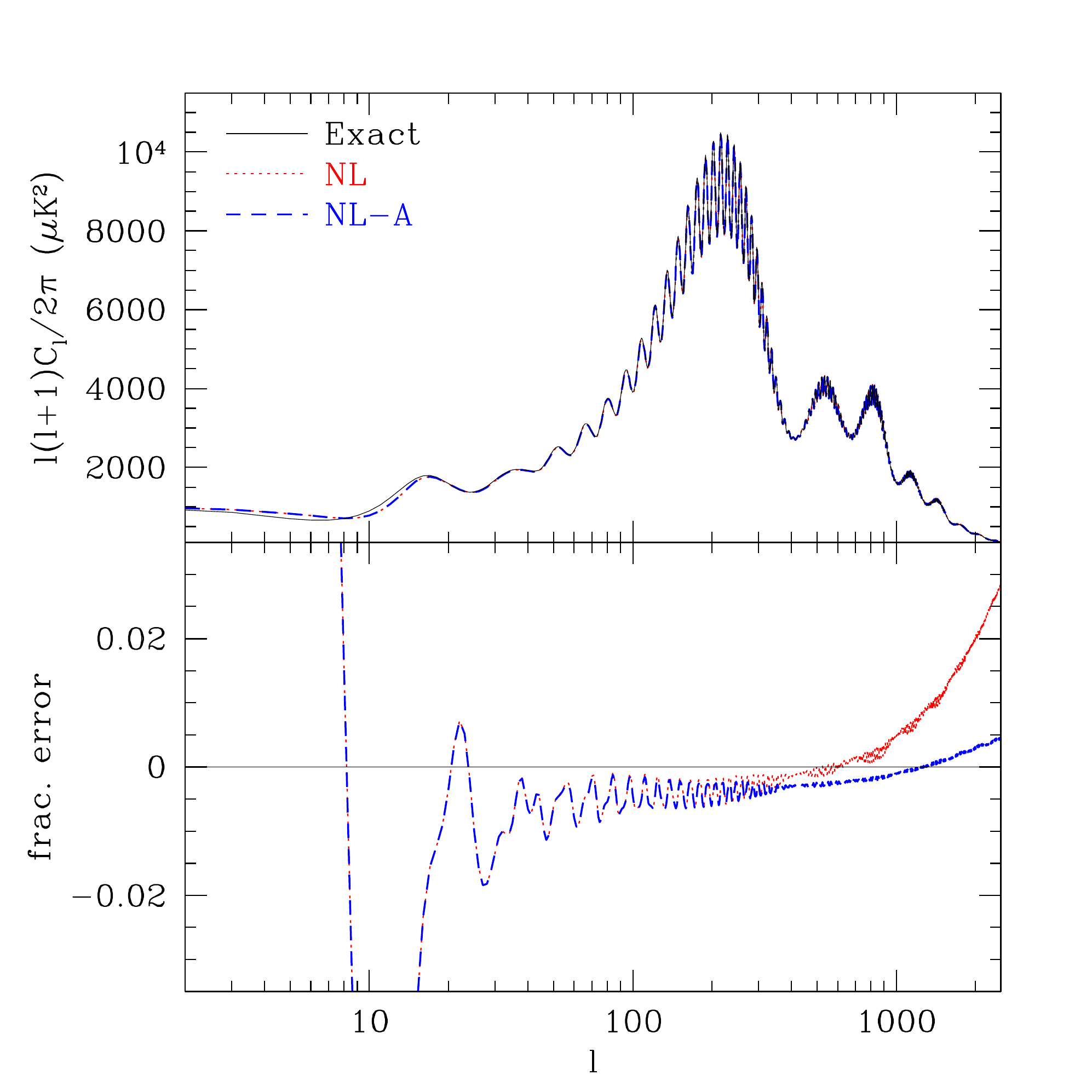, width=3.25in}
\caption{\footnotesize Curvature and CMB temperature power spectrum for the extremely large
step case of Fig.~\ref{plot:step_curv}.   The error in the damping of the oscillations can be further
reduced from the NL model (dotted) by adjusting the location and width of the feature
to $\eta_s \rightarrow 1.000029\eta_s$  and $x_d \rightarrow 0.936x_d$ with the same form (NL-A,  dashed).  Remaining errors
are comparable to or less than cosmic variance out to the CMB damping scale even for
this extremely large amplitude.}
 \label{plot:adjustment}	
\end{figure}

In fact most of the error in NL is not an error in the form of the template but rather a 
slight misassociation of the location $\eta_s$ and damping scale $x_s$ of the step given
potential parameters.    In Fig.~\ref{plot:adjustment} we show the result of adjusting these
parameters.   The main improvement is in the scale at which damping sets in due to
a $\sim 7\%$ decrease in $x_d$.   The small adjustment in $\eta_s$ makes the phase
a better match for $x< x_d$ with only a $3\times 10^{-5}$ change.  
For the analysis of the observational data, these small corrections can be applied
after the constraints on $6 b_V/\epsilon_{HB}$, $\eta_s$ and $x_d$ are obtained with the NL template.

Because damping makes a change in how efficiently excitations generate
further excitations we expect a change in the phase at $x>x_d$ that is not captured by the
NL form.   Since its effect comes in after the oscillations have already damped, this is a
small problem even for extremely large step.   If desired it can be fixed by introducing
a running to the phase offset $\bar\varphi(\ln k)$ in Eq.~(\ref{eqn:Bvarphi}).

 %------------------------------

\subsection{Monodromy}%%%%%%%%%%%%%%%%%%%%
\label{sec:monodromy}

Our second example is axion monodromy  where the potential is given by \cite{Silverstein:2008sg}
\begin{equation}
V(\phi) = \bar V(\phi) + \Lambda^4 \cos\left( \frac{\phi}{f_a}+\theta \right).
\end{equation}
We assume that $\Lambda^4/\bar V \ll \bar \epsilon_H$ so that the inflaton
can roll down the potential through the oscillations and that $f_a \ll 1$ in Planck units.
Here again $\bar V(\phi)$ is an underlying smooth slow-roll potential which in examples we take
as $\bar V=\lambda \phi$ so that $\bar\epsilon_H \approx 1/2\phi^2$.

\subsubsection{Source Model}%%%%%%%%%%

In order to calculate the excitations due to the oscillatory piece, we again need a model for the $f'/f$.
Following Refs.~\cite{Dvorkin:2009ne,Motohashi:2015hpa} we can in general approximate
\begin{equation}
G' + \frac{2}{3}\left(\frac{f'}{f}\right)^2 \approx 3\left( \frac{V_{,\phi}}{V} \right)^2 -2 \left( \frac{V_{,\phi\phi}}{V} \right),
\label{eqn:Gpmonodromy}
\end{equation}
up to $\epsilon_H$ suppressed corrections.  
Since $f_a \ll 1$, the $V_{,\phi\phi}$ term from the oscillations is dominant over the slow roll contributions
$G' \approx \delta G'$.
The $V_{,\phi}$ term
is suppressed compared with the second by ${\cal O}(\sqrt{\bar\epsilon_H} f_a)$ which we neglect.
Since the oscillatory part of the potential makes only transient
changes to the rolling of the inflaton we take~\cite{Flauger:2009ab}
\begin{equation}
\phi \approx \phi_* + \sqrt{ 2\bar\epsilon_H} \ln(x/x_*),
\label{eqn:phimodel}
\end{equation}
where $x_*$ is a suitably chosen normalization epoch which we optimize below (see \cite{Motohashi:2015hpa}).  
For sufficiently small $\Lambda^4/\bar V$ we can drop the quadratic term $(f'/f)^2$ in Eq.~(\ref{eqn:Gpmonodromy}) and integrate 
\begin{equation}
\delta G'=-2(\delta \ln f)' + \frac{2}{3}(\delta \ln f)''
\end{equation}
to define the oscillatory excitation source
\begin{equation}
\delta\ln f = - 3\frac{ \Lambda^4}{\bar V } 
\frac{\omega\cos(\omega\ln x+\psi)+ 3 \sin(\omega\ln x +\psi) }{f_a^2\omega (\omega^2 + 9)}  ,
\label{eqn:montoymodel}
\end{equation}
where the log frequency $\omega = \sqrt{2 \bar\epsilon_H}/f_a$ and
\begin{equation}
\psi = \frac{\phi_*}{f_a}- \omega \ln x_* + \theta.
\end{equation}
Since we are interested in subhorizon excitations we will hereafter assume $\omega>1$.
Dropping $(f'/f)^2$ in Eq.~(\ref{eqn:Gpmonodromy}) in this limit is equivalent to our original assumption that
$\Lambda^4/\bar V \ll \bar \epsilon_H$.

\subsubsection{Linear Excitations}%%%%%%%%%%

With these approximations, the first order integrals can be evaluated in closed form
\cite{Motohashi:2015hpa}
\begin{eqnarray}
\delta I_0 & \approx&  -A \cos(\bp-\psi) , \nonumber\\
\delta I_1 &\approx& -\frac{A}{\sqrt{2}} \tanh \left( \frac{\pi \omega}{2}\right) \sin(\bp-\psi)  ,
\label{eqn:monodromyI0}
\end{eqnarray}
where
\begin{eqnarray}
A  = \frac{3 \Lambda^4}{ \bar V \bar\epsilon_H \sqrt{ 1+ (3 /\omega)^2}} \sqrt{\frac{\pi\omega}{2} \coth\left( {\frac{\pi\omega}{2}}\right) },
\end{eqnarray}
and
\begin{eqnarray}
e^{i \bp} = -  2^{i\omega } \sqrt{ \frac{ \pi \omega}{ (9+ \omega^2)\sinh(\pi\omega)} }
\frac{(i+\omega )(3-i\omega )} {\Gamma(2+i\omega )}.
\end{eqnarray} 
For $\omega\gg 1$
\begin{eqnarray} 
\lim_{\omega\rightarrow \infty} (\delta \ln f)' & \approx & A\sqrt{\frac{\omega}{2\pi}} \sin(\omega\ln x+\psi) ,\nonumber\\
\lim_{\omega\rightarrow \infty} A &=& \frac{3 \Lambda^4}{\bar V \bar\epsilon_H}\sqrt{\frac{\pi \omega}{2}},\nonumber\\
\lim_{\omega\rightarrow \infty} \bp & = & \omega [ 1-\ln (\omega/2)]-\frac{\pi}{4},
\label{eqn:omegagg1}
\end{eqnarray}
where the amplitude $A$ and phase $\bp$ reflect the stationary phase evaluation for the integrals over the
oscillatory $(\delta \ln f)'$ (see below).  The condition $\Lambda^4/V \ll \bar \epsilon_H$ requires $A \ll \sqrt{\omega}$ but allows
greater than order unity effects in the power spectrum at sufficiently large frequency $\omega$.

Thus given the rapid convergence of $\tanh(\pi\omega/2)$ to unity for $\omega>1$, we can read off of
Eq.~(\ref{eqn:monodromyI0})
\begin{eqnarray}
B &\approx &  A + {\cal O}(A^3)  ,\nonumber\\
\delta \varphi &\approx & \bp-\psi + {\cal O}(A^2) .
\label{eqn:monleadingorder}
\end{eqnarray}
Our nonlinear ansatz of Eq.~(\ref{eqn:Bvarphi}) is equivalent to dropping the higher order corrections.
Similarly to the step model, if all of the excitations originated from the same resonance point
as the first order term, this truncation would provide an exact result according to 
Eq.~(\ref{eqn:recurse}).    Deviations from this approximation occur if the phase of the first order excitation
$\beta_1$ evolves due to excitation generating excitations  away from the resonance point.

\subsubsection{Nonlinear Excitations}%%%%%%%%%%

As with the step model, we can explicitly evaluate the \bogo hierarchy $\alpha_n$ and $\beta_n$ 
that define nonlinear excitations for monodromy.
In this case there are no simple closed form expressions beyond $n=1$ so before turning to numerical
results it is instructive to examine the parts of $\beta_1$ which then drive the higher excitations.
$\beta_1$ itself can be expressed in terms of hypergeometric functions but its main features can
be better understood in terms of contributions well before, at and well after resonance.

To simplify this treatment let us take the $\omega\gg 1$ limit of Eq.~(\ref{eqn:omegagg1}) 
and ignore slow roll evolution, which just renormalizes the coefficients, and set $\bar f = f_*=f_0$ so that 
\begin{eqnarray}
\beta_1(x) \approx -A\sqrt{\frac{\omega}{2\pi}}  \int_x^{\infty} \frac{du}{u} \sin(\omega\ln u+\psi) e^{2i u}.
\end{eqnarray}

Resonance occurs where the phase $p=2u \pm (\omega\ln u+\psi)$ reaches a stationary point
$d p/du =0$, namely at $u=\omega/2$~\cite{Flauger:2009ab}.     Using the stationary phase approximation for the integral,
we see that the impact of the resonance on $\beta_1$ is to give a step like resonant contribution
\begin{eqnarray}
\beta_1^{({\rm r})}(x) = \frac{A}{2} 
e^{i(\bp-\psi)} S,
\end{eqnarray}
where $S=0$ for $x \gg \omega/2$ and $1$ for $x \ll \omega/2$.   If this were the only contribution,
then like the step potential this would generate further excitations as
\begin{eqnarray}
\alpha_n^{({\rm r})}(x) &=& \frac{1}{n!} \left( \frac{A}{2} S \right)^n,   \quad (n={\rm even}), \nonumber\\
\beta_n^{({\rm r})}(x) &=&  e^{i(\bp-\psi)}\frac{1}{n!} \left( \frac{A}{2} S \right)^n , \quad (n={\rm odd}),
\label{eqn:resonantterms}
\end{eqnarray}
leading to the resummation
\begin{eqnarray}
\alpha^{({\rm r})}(x) &=& \cosh(A/2) , \nonumber\\
\beta^{({\rm r})}(x) &=& e^{i(\bp-\psi)}  \sinh(A/2) ,
\end{eqnarray}
after the resonance at $x \approx \omega/2$.  However $\beta_1$ also has transient oscillatory or nonresonant contributions well before and well after resonance
\begin{equation}
\beta_1^{({\rm nr})} \approx \frac{A}{2}\sqrt{\frac{\omega}{2\pi}}  e^{2 i x}
\left[ \frac{e^{-i(\omega\ln x+\psi)}}{2 x - \omega}  {-} \frac{  e^{i(\omega\ln x+\psi)}}{2 x + \omega}  \right],
\label{eqn:beta1nr}
\end{equation}
where $x \gg \omega/2$ or $x \ll \omega/2$.   This restriction is due to the fact that away from
the resonance the exponential factor oscillates rapidly and the slowly varying portion of the
integrand can be pulled out of the integral.   
While these contributions are transient and do not significantly impact the
freezeout value of $\beta_1$, they contain the same phase term $2 x \pm (\omega \ln x +\psi)$ as the modulated
source $(\delta \ln f)'e^{2ix}$ and can themselves generate further resonant excitations.   In effect, these
nonresonant terms in the first order excitation generate new resonances at higher order.  Since these
now occur for a wide range of $x$ away from $x=\omega/2$, they have a different phase from the first order  
contribution and therefore break the form of the general resummation in Eq.~(\ref{eqn:resummation}).

We can explicitly see this in its contribution to $\alpha_2$.  
There are terms from $\beta_1^{({\rm nr})}$ whose phase cancels or resonates with the modulated source
\begin{eqnarray}
\alpha_2^{({\rm nr})} &=&   i \frac{A^2 \omega}{2\pi} \int du \frac{1}{(2u-\omega)(2u+\omega)}+\ldots.  
\nonumber\\ 
&\sim &     i \frac{A^2}{4\pi}\int du \frac{1}{2u-\omega} +\ldots,
% \alpha_2^{({\rm nr})} &=&i \frac{A^2 \omega}{2\pi} \int_{u \not\in \omega/2} du \frac{1} { 4u^2-\omega^2}+\ldots,
% &&\quad \frac{ - \cos(2\omega\ln u+2\psi) +i \sin(2\omega\ln u+ 2\psi) \omega/2u}{\omega^2- 4u^2}
\end{eqnarray}
where in the second line we have approximated the integral around resonance.   
Note that this integral contains a divergent contribution at resonance due to the 
approximation in Eq.~(\ref{eqn:beta1nr}) that would be replaced by the actual resonant
terms of Eq.~(\ref{eqn:resonantterms}).   On the other hand, away from the resonance
the integral continues to contribute for a $\Delta u \sim \pm\omega$ leading to a net contribution
\begin{equation}
\alpha_2^{({\rm nr})} \sim  i {\cal O}(A^2 \ln \omega).
\end{equation}    
For sufficiently large frequency $\omega$ this contribution can dominate over those 
that are contemporaneous with the resonance in Eq.~(\ref{eqn:resonantterms}).    Due to cancellation
of the net effect on either side of the resonance, this term again is transient and does not survive to horizon
crossing.   However now $\beta_3$ has a source on either side of the resonance that is $\pi/2$ out of
phase with the resonant term.    The net effect is that the  phase $\varphi$ shifts 
due to nonlinear effects and to a lesser extent the amplitude of the oscillation increases
through these out of phase contributions to $B$.

Now let us quantify these considerations by evaluating the cubic term explicitly.
To capture the effects at horizon crossing and beyond we integrate the hierarchy in $\alpham$ rather than
$\alpha$ and $\beta$.   By direct computation using the full model for $\delta \ln f$ in Eq.~(\ref{eqn:montoymodel}), we obtain
\begin{eqnarray}
\lim_{x\rightarrow 0}\alpham_1(x) &=& \frac{1}{4} e^{i(\bp-\psi+\pi)} A , \nonumber\\
\lim_{x\rightarrow 0}\alpham_2(x) &=& \left[ \frac{1}{16} + {\cal O}(\omega^{-1})\right] A^2 , \nonumber\\
\lim_{x\rightarrow 0}\alpham_3(x) &=& C_3 e^{i(\bp-\psi + \phi_3)} {A^3}.
\end{eqnarray}
In $\alpham_2$ the $\omega^{-1}$ terms include contributions from the $I_2$ integral 
that break the nonlinear template form of Eq.~(\ref{eqn:template})
but are suppressed at high frequency.  We return to estimate these terms in the next section.

We can characterize the frequency dependence of the cubic amplitude and phase as
\begin{equation}
C_3 \approx 0.009804 \ln\omega + 0.02188,
\end{equation}
and
\begin{equation}
\phi_3/\pi \approx a \ln^2\omega + b\ln\omega + c,
\end{equation}
with $a= -5.40\times 10^{-4}$, $b=0.01356$, $c=-0.62359$ for $25 \lesssim \omega \lesssim 1000$.
The resonance prediction for these quantities from Eq.~(\ref{eqn:resonantterms}) would
be $C_3=1/96$ and $\phi_3=\pm \pi$.   Notice that the $c$ term means that
the phase shift is nearly $\pi/2$ as one might expect from the analytic arguments above.  
Furthermore the amplitude of the effect in $C_3$ grows logarithmically with $\omega$ making it 
the dominant effect at high frequency.
By matching the cubic expansion of the power spectrum template 
Eq.~(\ref{eqn:template}) to Eq.~(\ref{eqn:curviterative}), we obtain 
\begin{eqnarray}
B(A) &=&  A  -  \frac{1+96 C_3 \cos\phi_3}{24} A^3 + \ldots ,\nonumber\\
\delta\varphi(A) &=& \bp-\psi -4 C_3 \sin\phi_3 A^2 + \ldots,
\label{eqn:bvarphino}
\end{eqnarray}
which provides the next to leading order correction to Eq.~(\ref{eqn:monleadingorder}).
We use Eq.~(\ref{eqn:bvarphino}) in the template Eq.~(\ref{eqn:template}) as our nonlinear analytic approximation below.

\begin{figure}[t] %------------------------------ 
\psfig{file=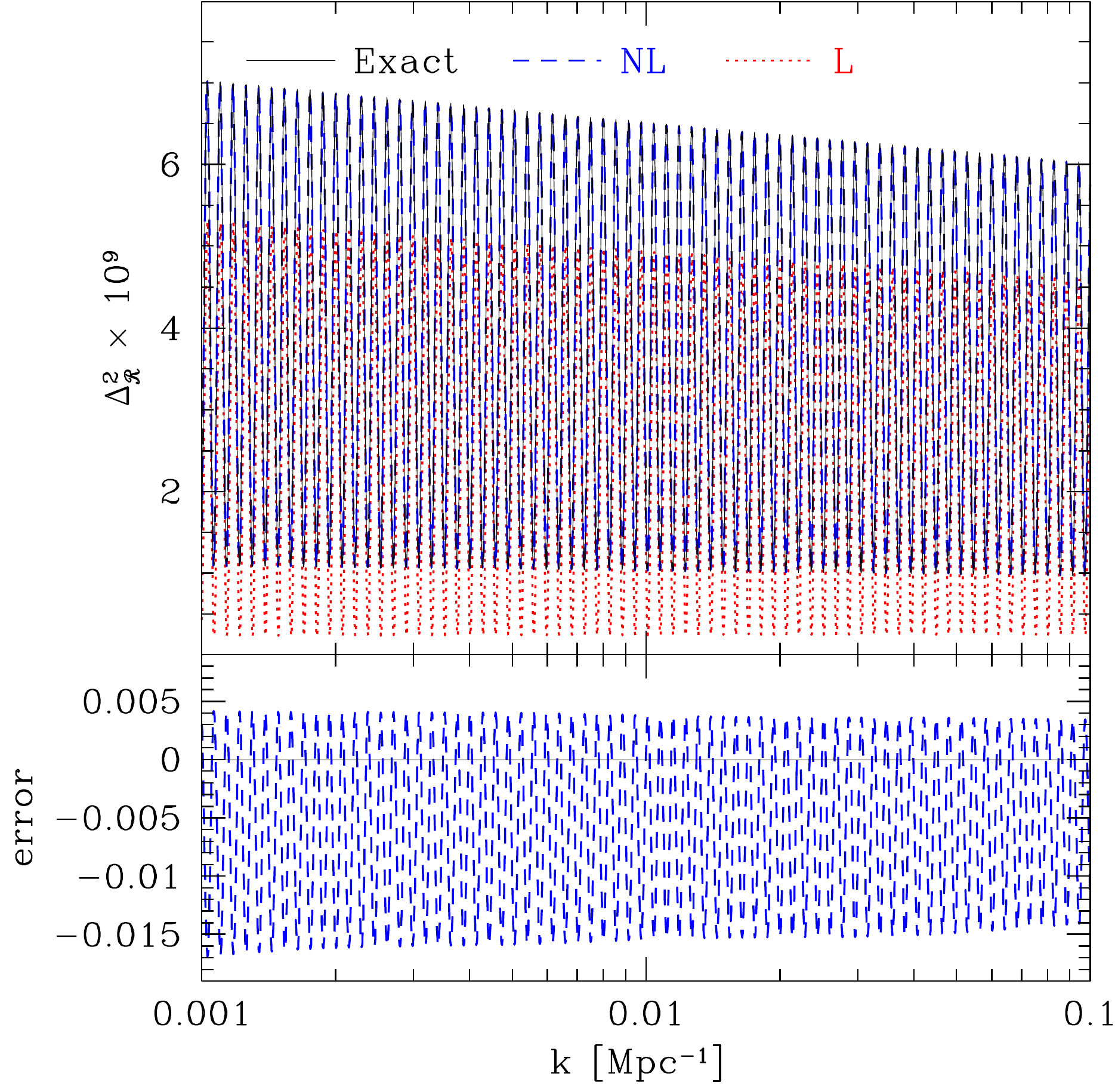, width=3.25in}
\psfig{file=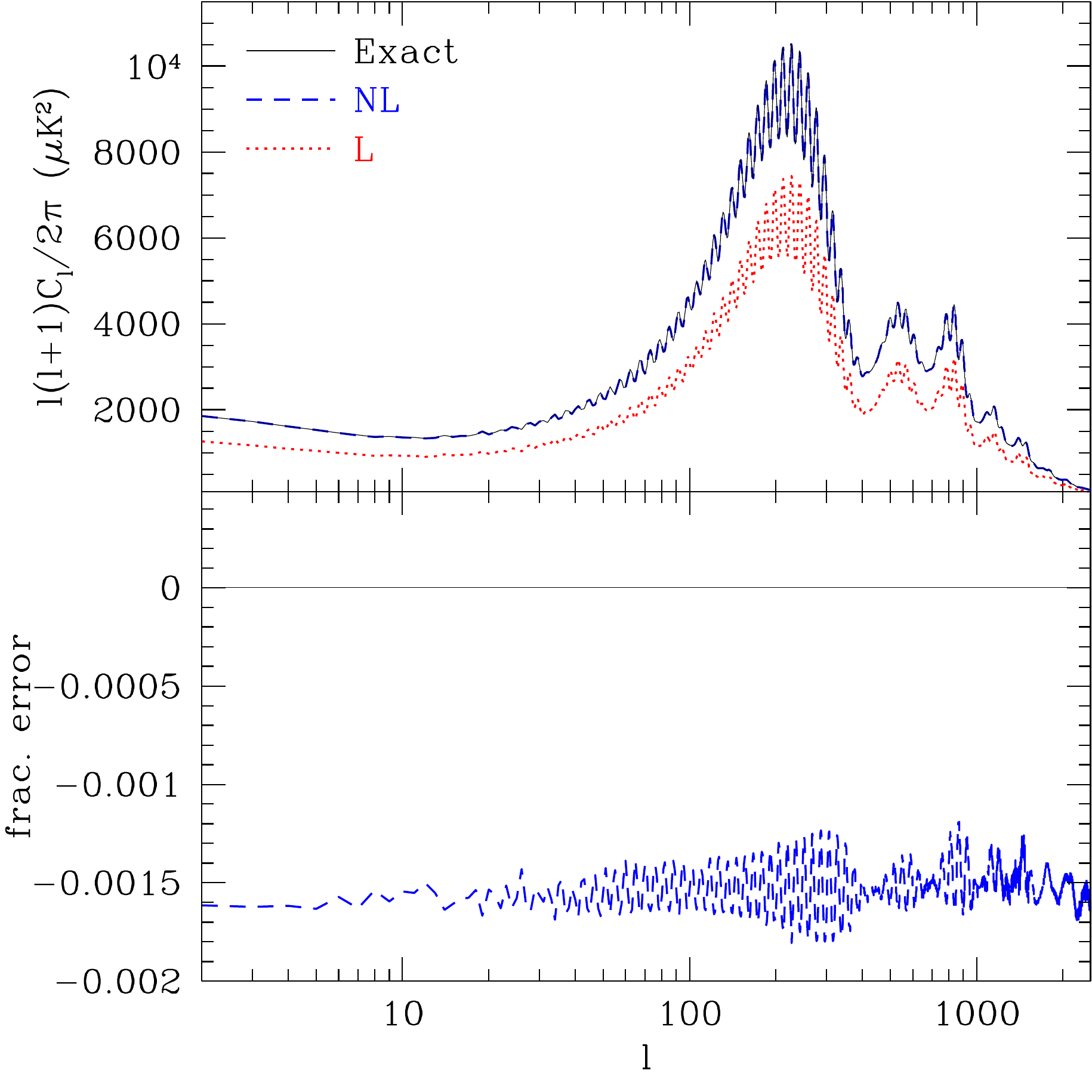, width=3.25in}
\caption{\footnotesize Monodromy curvature (upper) and CMB temperature (lower) power spectra
with  high frequency ($f_a^{-1}=1000$, $\omega \approx 100$), large amplitude ($A \approx 0.9$) oscillations.  Compared
are the numerical calculation (exact, solid), the analytic template prescription with nonlinear associations 
with potential parameters (NL,  dashed), and with the linearized associations (L, dotted).    L  errs in the
amplitude and zero point of oscillations (upper, offscale lower) whereas  NL  is accurate to $10^{-3}$ 
with a remaining error that can be mainly reabsorbed into
the slow roll amplitude.}
\label{plot:monodromyhigh}	
\end{figure} %------------------------------

\begin{figure}[t] %------------------------------ 
\psfig{file=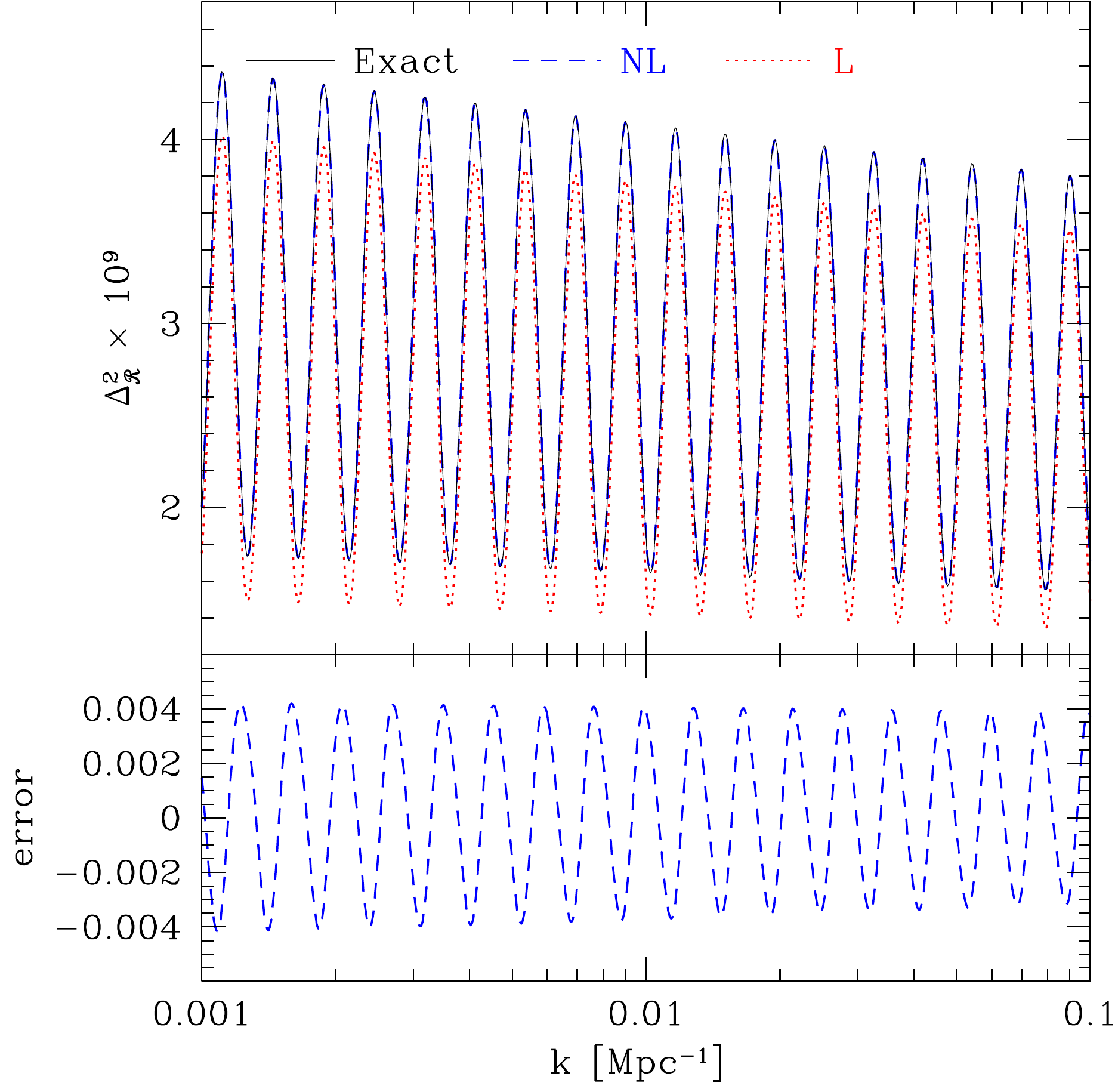, width=3.25in}
\psfig{file=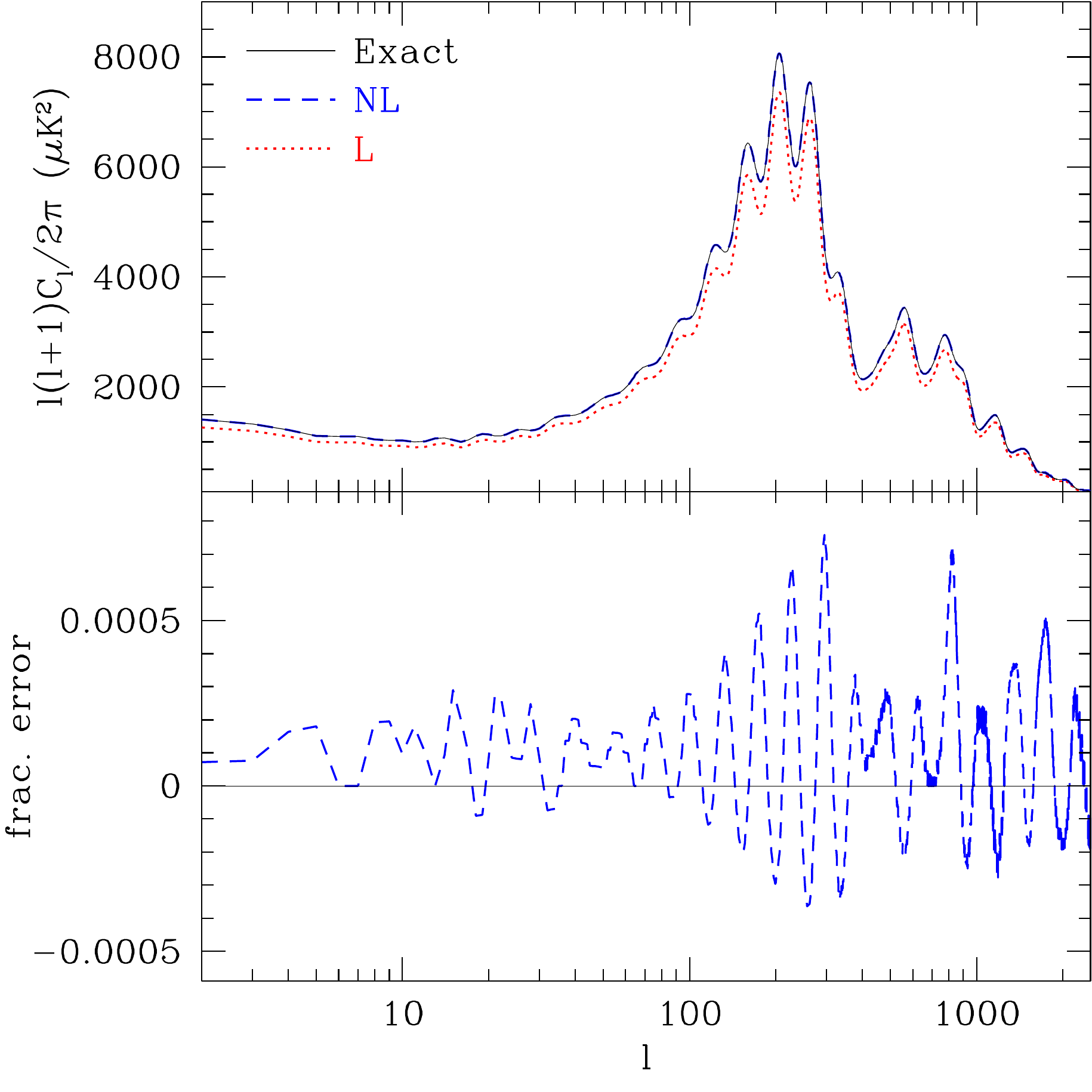, width=3.25in}
\caption{\footnotesize
Monodromy curvature (upper) and CMB temperature (lower) power spectra
with  moderately high frequency  ($f_a^{-1} = 250$, $\omega \approx 25$), large amplitude ($A\approx 0.45$) oscillations
(see Fig.~\ref{plot:monodromyhigh} for description).  Lower frequency oscillations suffer less from 
projection suppression and   allow only a smaller more linear $A$ for the same  deviations in the CMB which
are accurately captured by the NL prescription.}
\label{plot:monodromylow}	
\end{figure} %------------------------------

\subsubsection{Neglected Terms}%%%%%%%%%%

Eq.~(\ref{eqn:bvarphino}) neglects higher order terms in $B(A)$ and $\delta\varphi(A)$ as well as 
terms that are suppressed at high frequency that begin at quadratic order.  On the other hand through
$\sinh B$, the oscillation amplitude contains higher terms in  $A$ from $\alpham_4$ and higher.  
To check when higher order terms in $B(A)$ must be included compare the predictions
from Eq.~(\ref{eqn:bvarphino}) for
\begin{eqnarray}
\lim_{x\rightarrow 0}\alpham_4(x) &=& C_4 A^4, \nonumber\\
\lim_{x\rightarrow 0} \alpham_5(x) &=& C_5 e^{i(\bp-\psi + \phi_5)} {A^5},
\label{eqn:quarticquintic}
\end{eqnarray} 
to the direct computation of the integrals.
For example for $\omega=100$, the prediction for $C_4=0.00367$ whereas the integration gives
$0.00379$, whereas the prediction for $C_5=0.00875$ compared with $0.00870$ and
the prediction for $\phi_5=-0.149\pi$ compared with $-0.151\pi$.   Thus we expect the
truncation in  Eq.~(\ref{eqn:bvarphino}) to be a good approximation out to $A \sim {\rm few}$.  Since  
$\sinh B(A)$ exponentiates this quantity, this truncation should be valid out
to very high oscillation amplitude in the curvature spectrum.

We can also use this model to estimate the frequency suppressed terms from $I_2$ in $\alpham_2$
that break the form of the template constructed from $B$ and $\varphi$ in Eq.~(\ref{eqn:template}).
For $\omega \lesssim 100$, we can approximate $\delta\alpham_2=\alpham_2-{A^2}/16$ as
\begin{equation}
{\delta \alpham_2}\approx  C_{20} +C_{22} e^{2 i(\bp-\psi)+i \phi_2},
\end{equation}
where
\begin{eqnarray}
C_{20}  &\approx& -0.0802\omega^{-1} A^2 ,\nonumber\\
C_{22} &\approx&  0.0706\omega^{-3/2}A^2 ,\nonumber\\
\phi_2/\pi &\approx& -0.04 \omega + 0.69 .
\label{eqn:C2toy}
\end{eqnarray}
Since the model in Eq.~(\ref{eqn:montoymodel}) does not include all of the frequency suppressed contributions,
we use this calibration mainly to provide a template to monitor new terms in the power spectrum
\begin{eqnarray}
\label{eqn:templatebreak}
\Delta_\curv^2 &=&\bar\Delta_\curv^2 \Big[ 
\cosh B  - \sinh B \cos\varphi \\
&& + 4  C_{20}  + 4 C_{22} \cos(2\bp-2\psi+\phi_2)   \Big] . \nonumber
\end{eqnarray}
The quadratic nature of the sources in $I_2$ of Eq.~(\ref{eqn:I2}) produces oscillations with twice the frequency of the
resonant term as would other second order terms from low frequency or horizon scale effects (see \cite{Motohashi:2015hpa}).  
We use the form of Eq.~(\ref{eqn:templatebreak}) in the next section
to fit the results of the exact calculation  and monitor the neglected terms directly.

\subsubsection{Comparisons and Fits}%%%%%%%%%%

Before comparing our new nonlinear form of Eq.~(\ref{eqn:template}) and (\ref{eqn:bvarphino}), we review the common technique
for fitting the observed power spectrum to a  template based on the linearized analysis of Ref.~\cite{Flauger:2009ab,Ade:2015lrj}
\begin{equation}
\Delta_\curv^2 = A_s \left( \frac{k}{k_0} \right)^{n_s-1} \left(  1 - \delta n_s \cos \varphi \right),
\label{eqn:monotemplate}
\end{equation}
where $A_s$ $n_s$, $\delta n_s$ are taken to be constants and the phase is fit
to a logarithmic evolution around $\ln k_0$ \cite{Flauger:2014ana}
\begin{equation}
\varphi \approx \varphi_0 + \alpha_\omega\left( \ln \frac{k}{k_0} + \frac{c_1}{N} \ln^2 \frac{k}{k_0}+ \frac{c_2}{N^2}  \ln^3 \frac{k}{k_0}\right),
\end{equation}
where $N$ is a constant chosen to be of order the number of efolds to the end of inflation to normalize the
constants $c_1$ and $c_2$.    In our examples below we take $k_0=0.05$ Mpc$^{-1}$ which
is near the best constrained wavenumber for the Planck temperature data \cite{Miranda:2013wxa}.

The phenomenological template in Eq.~(\ref{eqn:monotemplate}) matches the {\it functional form} 
of the nonlinear template Eq.~(\ref{eqn:template}) for $\delta n_s < 1$,  with the associations (NL)
\begin{eqnarray}
A_s &=& \bar\Delta_\curv^2 (k_0)\cosh B, \quad
\delta n_s = \tanh B,   \nonumber\\
\label{eqn:correspondNL}
\varphi_0  &=& \varphi(k_0), \quad
\alpha_\omega =  \frac{d\varphi}{d\ln k}\Big|_{k_0}  , \\
\frac{c_1}{N}\alpha_\omega &=& \frac{1}{2} \frac{d^2\varphi}{d\ln k^2}\Big|_{k_0} , \quad
\frac{c_2}{N^2}\alpha_\omega = \frac{1}{6}  \frac{d^3\varphi}{d\ln k^3}\Big|_{k_0}, \nonumber
\end{eqnarray}
but these differ from the linearized relations (L)
\begin{eqnarray}
A_s &=& \bar\Delta_\curv^2 (k_0), \quad
\delta n_s = A, \nonumber\\
\label{eqn:correspondL}
\varphi_0 & =& (\bp-\psi)\Big|_{k_0}, \quad
\alpha_\omega =  \frac{d(\bp-\psi)}{d\ln k}\Big|_{k_0} , \\
\frac{c_1}{N}\alpha_\omega &=& \frac{1}{2} \frac{d^2(\bp-\psi)}{d\ln k^2}\Big|_{k_0} , \quad
\frac{c_2}{N^2}\alpha_\omega = \frac{1}{6}  \frac{d^3(\bp-\psi)}{d\ln k^3}\Big|_{k_0}, \nonumber
\end{eqnarray}
which would be used to interpret the constraints  as $\delta n_s$ approaches unity.
Note that the linearized calculation would associate
power from the excitations themselves $\bar\Delta_\curv^2 \cosh B$ with the
underlying scalar amplitude of the slow roll potential $A_s$.   Since high frequency potential features
largely do not effect tensor modes \cite{Hu:2014hoa}, this leads to incorrect inferences about the 
scalar-tensor ratio and its consistency with the tensor tilt.

For example Ref.~\cite{Ade:2015lrj} restricts searches for monodromy oscillations in the
range to $\delta n_s <0.7$ and $1 < \omega < 10^3$.  
In this regime, the nonlinear corrections introduced here are $<40\%$ in $A_s$, $<24\%$ in $\delta n_s$ and $<0.27$ in the phase.   
Our NL correspondence would correct these errors in interpretation and allow searches to higher amplitude.

\begin{figure}[t] %------------------------------ 
\psfig{file=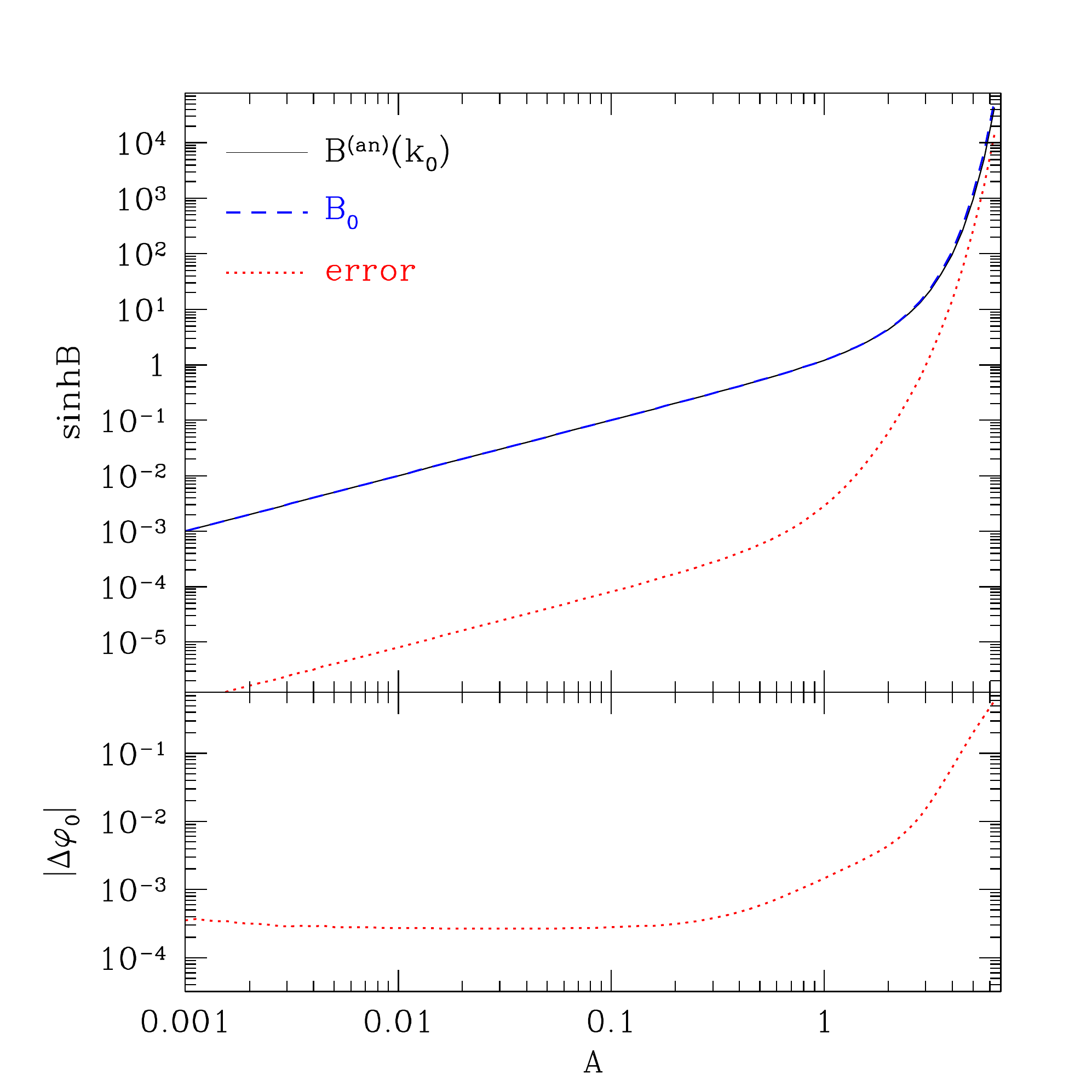, width=3.5in}
\caption{\footnotesize 
Analytic vs.~fitted template parameters $B$ and $\varphi$ at $k_0$ for an extremely high
frequency monodromy case ($f_a^{-1}=2500$, $\omega\approx 250$).    
Shown are the power spectrum
oscillation amplitude $\sinh B$ for the analytic (upper, solid) and fitted (upper, dashed) results, 
their absolute difference (upper, dotted) and the phase difference (lower, dotted).  
The analytic form reproduces the oscillation amplitude, zero point, and phase to a fraction of their values 
even when the former reaches $10^3$ times the slow roll power spectrum. }
\label{plot:templatehigh}
\end{figure} %------------------------------

\begin{figure}[t] %------------------------------
\psfig{file=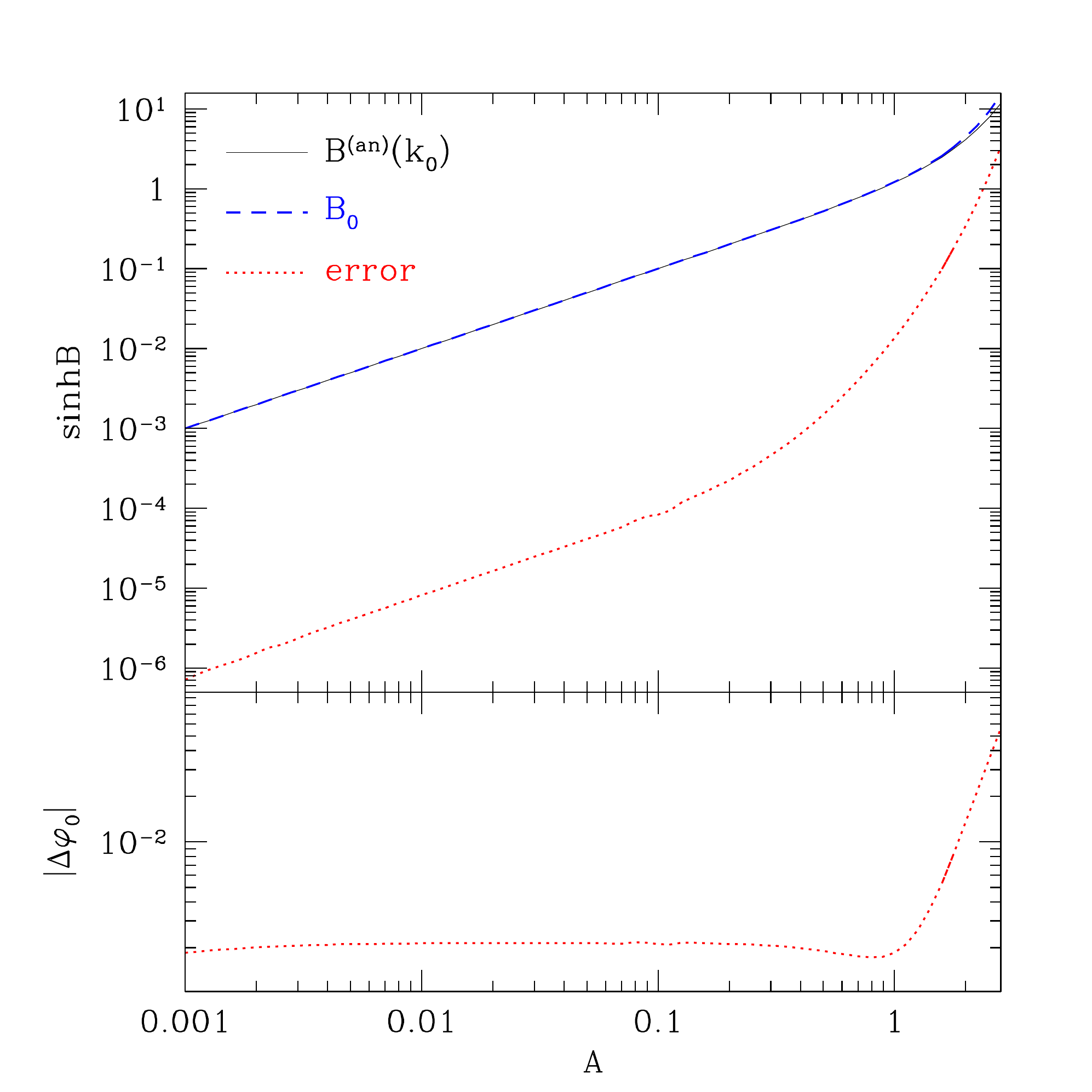, width=3.5in}
\caption{\footnotesize 
Analytic vs fitted template parameters $B$ and $\varphi$ at $k_0$ for the moderately high
frequency monodromy case ($f^{-1}=250$, $\omega\approx 25$).  Curves are same as in
Fig.~\ref{plot:templatehigh} with agreement extending to oscillation amplitudes that are 10 times the
slow roll power spectrum.}
\label{plot:templatelow}
\end{figure} %------------------------------

We now compare the analytic predictions with the exact calculation for example cases.
In order to better separate out small effects that would be captured by the running of slow roll parameters
from the new effects associated with nonlinear excitations, we evaluate
the parameters of the template forms as follows.   
While Eq.~(\ref{eqn:phimodel}) for the field position does not include running of slow roll parameters, we
can enhance the accuracy of these calculations by choosing the normalization point $\phi_*$ as well as
the calculation of $\omega$ and $A$ separately for each mode.    
As discussed in Ref.~\cite{Motohashi:2015hpa}, for $\omega>1$ the optimal point is the resonance point where 
$x_*=k\eta(\bar\phi)=\omega(\bar\phi)/2$.    We identify the resonance
point by solving this equation where in practice rather we calculate $\bar\phi(\ln \eta)$ by convolving
the field position $\phi(\ln \eta)$ with a Gaussian of width  of order the resonance
$\Delta \ln \eta = \sqrt{2/\omega}$ in order to remove the small oscillatory effects.

Likewise, when computing $\bar\Delta_\curv^2(k)$, we use the exact solution on the
$A=0$, $V=\bar V$ slow roll potential at the same field value at freezeout for each $k$-mode.    Finally we use the full
evaluation of $\varphi$ or $\bp-\psi$ in the analytic formulae rather than the Taylor expansions in
Eq.~(\ref{eqn:correspondNL}), (\ref{eqn:correspondL}) around $k_0=0.05$ Mpc$^{-1}$.   Thus deviations of the predictions 
from the exact calculation even at the level of the currently observationally negligible
${\cal O}(n_s-1)^2$ can be attributed to effects from the excitations rather than slow-roll evolution.

In Fig.~\ref{plot:monodromyhigh} we show an example with a very high frequency
$f_a^{-1} =1000$ or $\omega \approx 100$ and amplitude $A \approx 0.9$.  
Here and below we choose $\lambda=2.9609 \times 10^{-10}$ and $\theta=0.59726$ for definiteness.
Using the NL template form, the oscillation amplitude $\delta n_s=0.723$ and is near the edge of the region searched in
Ref.~\cite{Ade:2015lrj}.   Note that the NL template parameters predict the oscillation 
amplitude, phase and zero point in the curvature power spectrum to high accuracy (upper panel).  
Using the linearized form errs in all three quantities as expected even after removing slow roll drifts as described above.  
In the CMB temperature power spectrum, the oscillation amplitude is reduced by
projection effects leaving the change in the zero point or effective scalar amplitude
even more apparent (lower panel).   Fractional errors in the oscillatory part for the NL
template are comparable to or smaller than the cosmic variance limit all the way through the CMB damping tail.  
In fact we shall see that the main source of error can be removed
by making small adjustments to the template parameters $B$ and $\varphi$ that merely
change their association with the underlying potential amplitude $\Lambda^4$ and
phase $\theta$ by a comparable amount.

\begin{figure}[t] %------------------------------
\psfig{file=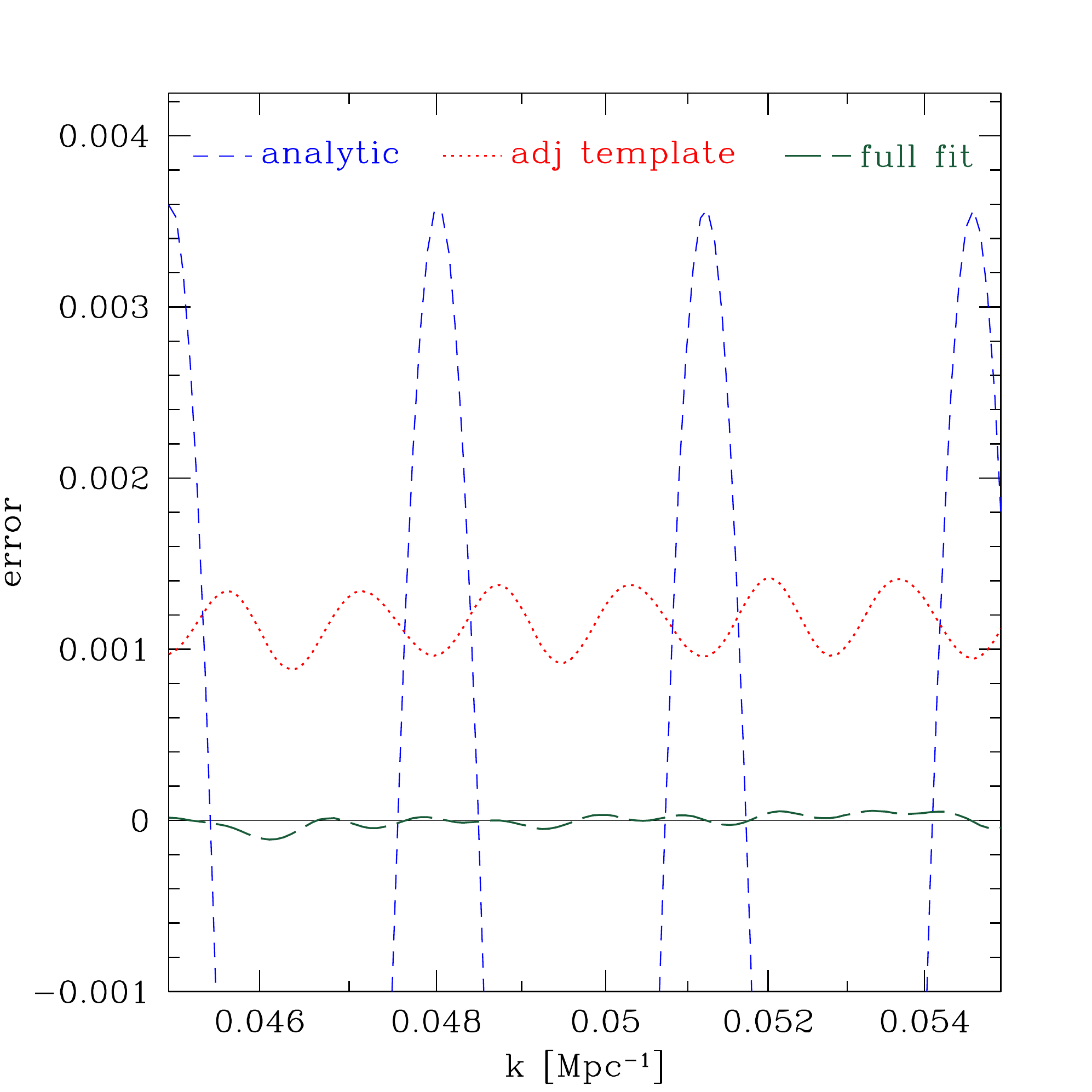, width=3.25in}
\caption{\footnotesize
Template parameter adjustment and template breaking terms for
the  high frequency $\omega \approx 100$ monodromy model of Fig.~\ref{plot:monodromyhigh}. 
Shown are the $\Delta_\curv^2$ errors vs the exact computation with the analytic template parameters $B,\varphi$ (dashed),  
best fit template parameters (dotted), and the full fit including $C_{20}, C_{22}, \phi_2$ (long dashed).  
Most of the error is removed by the template readjustment which uncovers a much smaller offset, double
frequency component that breaks the form of the high frequency template.   
The error after fitting this term is reduced to an entirely negligible level.}
\label{plot:refithigh}
\end{figure} %------------------------------

\begin{figure}[t] %------------------------------ 
\psfig{file=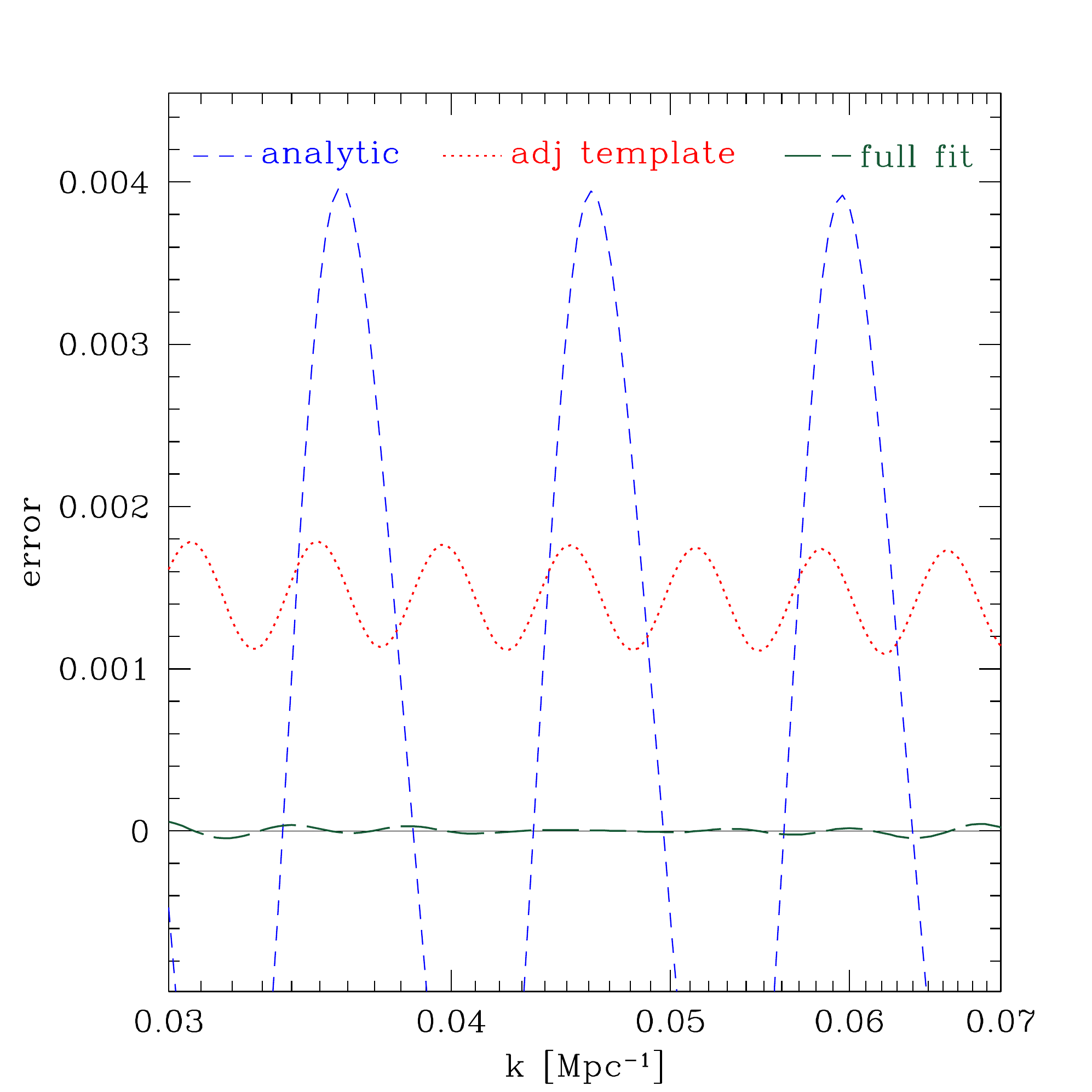, width=3.25in}
\caption{\footnotesize Template parameter adjustment and template breaking terms for
the moderately high frequency $\omega \approx 25$ monodromy model of Fig.~\ref{plot:monodromylow}.   
Curves are the same as in Fig.~\ref{plot:refithigh}.}
\label{plot:refitlow}
\end{figure} %------------------------------

In Fig.~\ref{plot:monodromylow} we show an example with a more moderate frequency $f_a^{-1}=250$ or
$\omega \approx 25$ and a smaller amplitude $A \approx 0.45$.  The smaller amplitude
is chosen both because the smaller projection effects make oscillations in the CMB 
temperature power spectrum more prominent and because in our approximations
we assume $A/\sqrt{\omega} \ll 1$.    For this lower amplitude, we achieve comparable precision in both the 
curvature and temperature power spectrum as in Fig.~\ref{plot:monodromyhigh}.

To further test the accuracy of the NL template approximation we can fit for the oscillation
amplitude, phase and zero point across a wide range in $A$ and $\omega$ and compare them with the predictions.   
First we consider $B$, $\varphi$, to be fit parameters at each $A$, $\omega$ for the exact power spectrum near $k_0$.  
In order to better capture the effects of the evolution of these
quantities away from $k_0$ across several oscillations, we use the close agreement with
the analytic forms to fit $B_0$, $\Delta\varphi_0$ and $\Delta\alpha_\omega$ as
\begin{eqnarray}
B(k) &=& B_0 \frac{B^{(\rm an)}(k)}{B^{(\rm an)}(k_0)}, \nonumber\\
\varphi(k) &=& \varphi^{(\rm an)}(k)+ \Delta\varphi_0 + \Delta\alpha_\omega \ln (k/k_0).
\end{eqnarray}
We also fit  the terms $C_{20}$, $C_{22}$ and $\phi_2$ in Eq.~(\ref{eqn:templatebreak}) to quantify the
frequency suppressed terms that break the nonlinear template.   Note that though $C_{20}$ 
violates Eq.~(\ref{eqn:template}), it can be reabsorbed into a change in $A_s$ and $\delta n_s$ in
Eq.~(\ref{eqn:monotemplate}) whereas $C_{22}$ would break either template.

\begin{figure}[t] %------------------------------ 
\psfig{file=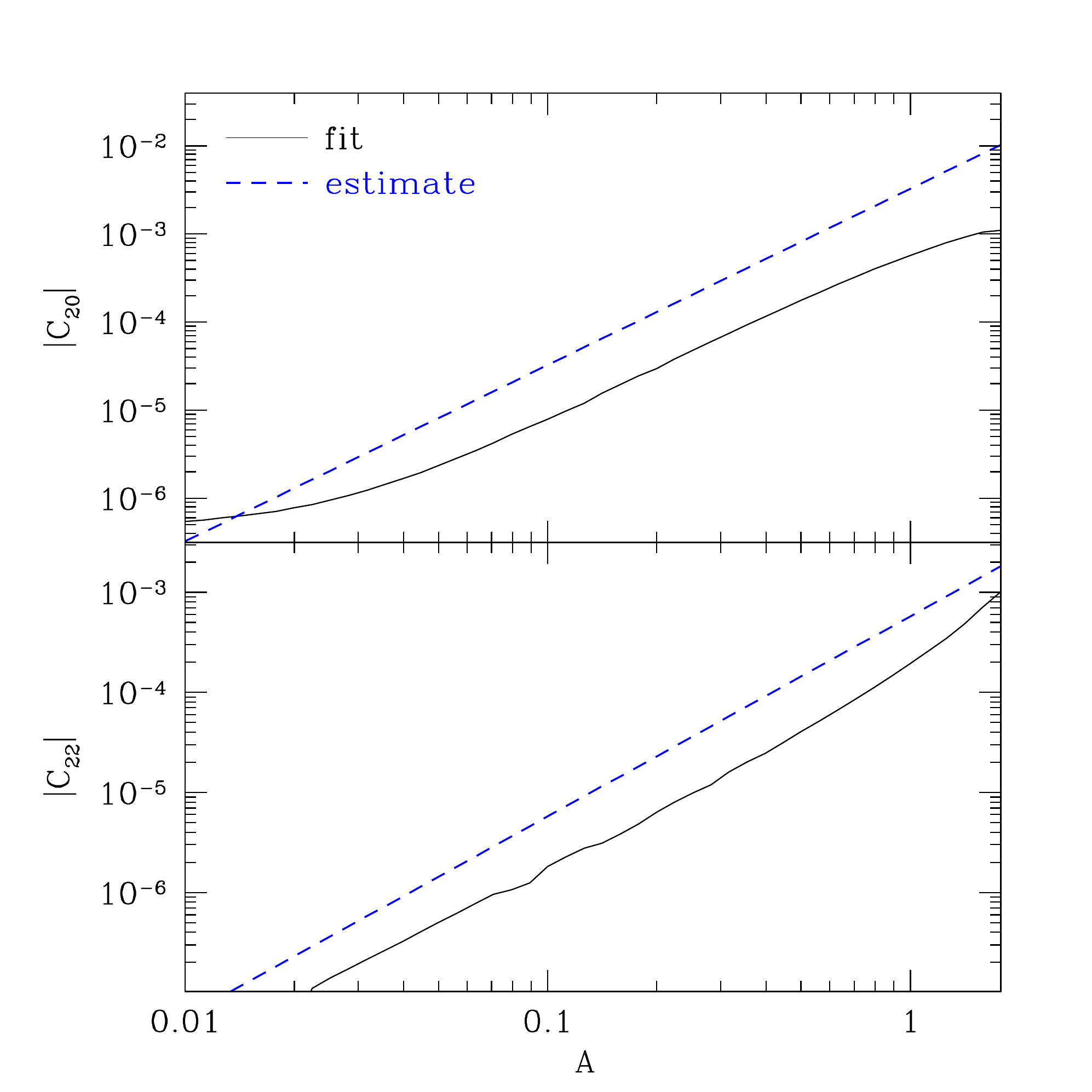, width=3.25in}
\caption{\footnotesize Template breaking offset double frequency components for the moderately high
frequency $(\omega\approx 25)$ monodromy case.  The estimates of Eq.~(\ref{eqn:C2toy}) are
$\sim 4$ times too high but otherwise usefully capture the scaling with $A$ and $\omega$.}
\label{plot:doubleparams}
\end{figure} %------------------------------

In Fig.~\ref{plot:templatehigh}, we show the result for an extremely high frequency 
($f_a^{-1}=2500$, $\omega\approx 250$).   This is near the highest frequency that is under control
in the effective theory that underlies the calculation \cite{Behbahani:2011it,Adshead:2014sga}.  
Remarkably, even when the oscillation amplitude
reaches $\sinh B \approx 10^3$, or a thousand times the slow roll power spectrum, the analytic
approximation for the amplitude and phase are accurate to a fraction of their values.  
This is consistent with the estimate from the quartic and quintic terms in Eq.~(\ref{eqn:quarticquintic}).

In Fig.~\ref{plot:templatelow}, we show the same comparison for the moderately high frequency
($\omega\approx 25$) case.   Due to neglected terms at low frequency in the model of Eq.~(\ref{eqn:montoymodel}), 
the good agreement extends out to a smaller $\sinh B\approx 10$ which is still 
an order of magnitude larger oscillation amplitude than the slow roll power spectrum.

These fits also uncover the offset double frequency terms  in Eq.~(\ref{eqn:templatebreak}).   
In Fig.~\ref{plot:refithigh}, we show the breakdown of the errors in the analytic $\Delta_\curv^2$ approximation 
in the high frequency model of Fig.~\ref{plot:monodromyhigh}.
Once $B$ and $\varphi$ are adjusted with the fitted values of $B_0$, $\Delta \phi_0$ and $\delta\alpha_\omega$, the
remaining errors are reduced by an order of magnitude and are of the form of an offset
double frequency component.  These fit well to the functional form given by $C_{20}$, $C_{22}$ and $\phi_2$ 
leaving the errors in the full fit at a negligible level.

These terms are not part of the high frequency template (\ref{eqn:template}) and their
contributions increase at lower frequency for a fixed $A$ at roughly the rate expected from 
Eq.~(\ref{eqn:C2toy}).  However due to projection effects, the same level of CMB temperature fluctuations
requires a smaller $A$.   For the moderate frequency and amplitude model of Fig.~\ref{plot:monodromylow}, the template breaking 
effects are at a comparable level (cf.~Figs.~\ref{plot:refithigh},\ref{plot:refitlow}).   
More generally we expect that for observationally viable models these terms which are neglected in 
the template form of fits to CMB data should make little to no impact on current constraints.

The rough estimate of Eq.~(\ref{eqn:C2toy}) can provide a useful guide for deciding when in the
future these terms would need to be considered.   In Fig.~\ref{plot:doubleparams}, we compare
these estimates to the fitted values of $C_{20}$ and $C_{22}$ for the moderately high $(\omega\approx25)$
case.   Eq.~(\ref{eqn:C2toy}) overestimates these effects by roughly a factor of 4 but otherwise
usefully captures the scaling  for $0.03 \lesssim A \lesssim 1$ and few $\lesssim \omega
\lesssim 100$.   For $\omega \lesssim 1$ the fully analytic results from Ref.~\cite{Motohashi:2015hpa}
can instead be applied to quantify accurately these double frequency terms.

\section{Discussion}%%%%%%%%%%%%%%%%%%%%%%%%%%%%%%%%%%%%%%%
\label{sec:discussion}

In this work we have introduced a new formalism to calculate features in the curvature spectrum  
to arbitrary order in their deviation from the scale-free slow roll form.  
It improves on previous techniques involving the inflaton modefunction~\cite{Stewart2002} 
by allowing a straightforward iteration that conserves curvature fluctuations order by order.  
This is important for models with order unity curvature features and larger
where the iterative series must be resummed into a nonlinear form or 
where precision measurements require higher than second order accuracy.

Using these techniques, we show that for high frequency, or subhorizon, excitation of
the curvature modefunction by features, the relationship between \bogo excitation 
coefficients restricts the nonlinear form of power spectrum features.    Due to this
relationship, the first nontrivial effect of subhorizon excitations generating further excitations 
arises at third order.   We show that if this process occurs contemporaneously 
with the original excitation, then the series can be resummed in closed form and is 
a simple exponentiation of the linearized response.  
Even for cases where this exact relationship does not hold, this exponentiated form provides
a physically motivated and controlled template for fitting features in the power spectrum.  
It furthermore exactly matches second order perturbation theory in the small feature, high frequency limit.  
It improves on a similar exponentiation ansatz in Ref.~\cite{Dvorkin:2009ne} by enforcing the nonlinear
relationship between the \bogo excitation coefficients.

Applied to the step and axion monodromy models, these techniques greatly improve
the accuracy of predictions for curvature and CMB temperature power spectrum features
directly from potential parameters.    In the step case, the improved form of the template
allows greater than order unity curvature oscillations to be fit to better than cosmic variance
accuracy for CMB measurements out through the damping tail.    For monodromy, the
improvement is mainly in the mapping between potential parameters and phenomenological
parameters that describe the amplitude, phase and zero point of the logarithmic oscillations.
Remarkably, our  analytic description  reproduces all three to good approximation even
when the oscillation amplitude is $10^3$ times the slow roll power spectrum for models with sufficiently high frequency.  
Of course, this relates only to the formal accuracy of solutions to the Mukhanov-Sasaki equation
(\ref{eqn:Reqn}), whereas correct predictions also require the validity of the  effective theory that underlies it.
We also estimate when terms that produce double frequency oscillations,
which are absent in the subhorizon template, should be included when analyzing data.

These techniques will enable future studies of CMB and large scale
structure power spectra to extend to high amplitude, high frequency features in specific cases such as steps
and axion monodromy as well as in model-independent searches for temporal features
during inflation.

\acknowledgements%%%%%%%%%%%%%%%%%%%%%%%%%%%%%%%%%%%%%%%

We thank P.~Adshead, R.~Flauger, A.~Joyce for useful conversations.   
This work was supported by U.S.~Dept.\ of Energy contract DE-FG02-13ER41958, NASA ATP NNX15AK22G and by
the Kavli Institute for Cosmological Physics at the University of Chicago through grants NSF PHY-1125897 
and an endowment from the Kavli Foundation and its founder Fred Kavli.  
VM was supported in part by the Charles E.~Kaufman Foundation, a supporting organization of the Pittsburgh Foundation.
WH thanks the Aspen Center for Physics, which is supported by National Science Foundation grant PHY-1066293, 
where part of this work was completed.

\bibliography{nonlinex}
\end{document}